\documentclass[aps, showkeys]{revtex4}

\usepackage[small]{caption}
\usepackage{amsmath,amsfonts,amscd,amssymb,epsf, epsfig,mathrsfs,graphicx,color}

\begin{document}

\title{Finite temperature Casimir effect on  spherical shells in $\boldsymbol{(D+1)}$-dimensional spacetime and its  high temperature limit}
\author{L.P. Teo}\email{ LeePeng.Teo@nottingham.edu.my}\address{Department of Applied Mathematics, Faculty of Engineering, University of Nottingham Malaysia Campus, Jalan Broga, 43500, Semenyih, Selangor Darul Ehsan, Malaysia. }

\begin{abstract}
We consider the finite temperature Casimir free energy acting on a   spherical shell in $(D+1)$-dimensional Minkowski spacetime due to the vacuum fluctuations of scalar and electromagnetic fields. Dirichlet, Neumann, perfectly conducting and infinitely permeable boundary conditions are considered. The Casimir free energy is regularized using zeta functional regularization technique. To renormalize the Casimir free energy, we compute the heat kernel coefficients $c_n$, $0\leq n\leq D+1$, from the zeta function $\zeta(s)$. After renormalization, the high temperature leading term of the Casimir free energy is $- c_DT\ln T-T  \zeta'(0)/2$. Explicit expressions for the renormalized Casimir free energy  and $  \zeta'(0)$ are derived. The dependence of the renormalized Casimir free energy on temperature is shown graphically.
\end{abstract}

 \maketitle

 \section{Introduction}

Casimir effect is an interesting quantum phenomenon that has attracted the attention of physicists for more than 60 years \cite{9}.  In recent years, the finite temperature correction to the Casimir effect has attracted increasing interest both theoretically and experimentally \cite{9,46,49,50,51}. It has been known that in the high temperature regime, the vacuum fluctuations of quantum fields can give  rise to Casimir effect that is very different from the effect at zero temperature.

Since the pioneering work by
Ambj\o rn and Wolfram \cite{2}, where   Casimir energies for   scalar field and electromagnetic field in a $D$-dimensional rectangular
cavity were calculated, Casimir effect in higher dimensional spacetime has gradually received  more attention, especially in recent years, when some long standing problems in physics have found solutions by embedding our physical $(3+1)$-dimensional spacetime in higher dimensional spacetime.

Calculations of Casimir energy in spherically symmetric configurations have been of interest to physicists for more than 45 years now.
In \cite{17}, Boyer calculated the Casimir force acting on a perfectly conducting three dimensional spherical shell and found that it is repulsive. This result was later confirmed by several groups of researchers \cite{18,16,20,21,22,23,24,25} using different  methods.   The Casimir energy of a massive scalar field confined in the interior and exterior of a three dimensional spherical shell  was computed in \cite{26}.  Generalization to   higher dimensions was   carried out in \cite{27,28,29,30}.  Later in \cite{31}, a systematic approach for  scalar field, spinor field and electromagnetic field in $D$-dimensional spherical cavity was given. In \cite{10}, we reconsider the electromagnetic Casimir effect of a $D$-dimensional spherical shell. Most of these works only considered the zero temperature Casimir effect. So far finite temperature Casimir effect on spherical shells has only been considered for the case when $D=3$   \cite{9,16}.

In this work, we consider the finite temperature Casimir effect for higher dimensional spherical shells. We treat scalar field with Dirichlet and Neumann boundary conditions, and electromagnetic field with perfectly conducting and infinitely permeable boundary conditions. For scalar field, the results for zero temperature Casimir energy of a $D$-dimensional spherical shell can be found in   \cite{31, 9, 5}, whereas the corresponding results for electromagnetic field were obtained in \cite{10}.  The current work can be considered as an extension of these to the finite temperature region. Of particular interest to us is the high temperature asymptotic behavior of the Casimir free energy.

In this paper, we use units where $\hbar=c=k_B=1$.
\section{ Casimir free energy of a  spherical shell in $\boldsymbol{(D+1)}$-dimensional Minkowski spacetime}\label{sec3}

The Casimir free energy of a quantum field in a $D$-dimensional bounded region $M$ is defined as
\begin{align}
\label{eq10_17_3}E_{\text{Cas}}(M) =\frac{1}{2}\sum_{\omega > 0}\omega +T\sum_{\omega > 0}\ln\left(1-e^{-\omega /T}\right),
\end{align}where $\omega$ are the eigenfrequencies of the field.
 The zero temperature Casimir energy
\begin{align}
\label{eq10_17_4}E_{\text{Cas}}^{T=0}(M) =\frac{1}{2}\sum_{\omega > 0}\omega
\end{align}
 is generically divergent, but the thermal correction
 $$\Delta_TE_{\text{Cas}}(M) = T\sum_{\omega > 0}\ln\left(1-e^{-\omega /T}\right) $$is always finite.
Using zeta regularization, the regularized Casimir free energy is defined as (see e.g. \cite{5,9,11}):
\begin{equation}\label{eq10_17_8}
E_{\text{Cas}}^{\text{reg}}(M) =-\frac{T}{2}\left(\zeta_{T}'(M;0)+\ln [\mu ]^2\zeta_{T}(M;0)\right),
\end{equation} where $\mu$ is a normalization constant, and
\begin{align*}
\zeta_T(M;s)=\sum_{\omega>0}\sum_{p=-\infty}^{\infty}\left(\omega^2+[2\pi pT]^2\right)^{-s}
\end{align*}is the corresponding thermal zeta function. Let
\begin{align*}
 K (M;t)=\sum_{\omega> 0}e^{-t\omega^2}
\end{align*}be the associated heat kernel.
It is well known that the heat kernel has an asymptotic expansion of the following form as $t\rightarrow 0^+$ (see e.g. \cite{5,9,11}):
\begin{equation}\label{eq10_17_10}
K(M;t)\sim \sum_{n=0}^{\infty} c_{n}(M)t^{\frac{n-D}{2}}.
\end{equation}The coefficients $c_{n}$ are related to the zeta function
\begin{equation*}
\zeta(M;s)=\sum_{\omega> 0} \omega^{-2s}
\end{equation*} by
\begin{equation*}
c_{n}(M)=\text{Res}_{s=\frac{D-n}{2}}\left(\Gamma(s)\zeta(M;s)\right).
\end{equation*}
It has been shown that in the high temperature region (see e.g. \cite{9,6,15}), the regularized Casimir energy behaves as
\begin{equation}\label{eq10_17_12}
\begin{split}
  E_{\text{Cas}}^{\text{reg}}(M) \sim &-\frac{1}{\sqrt{\pi}}\sum_{n=0}^{D-1}2^{D-n} \Gamma\left(\frac{D-n+1}{2}\right)\zeta_R(D-n+1) c_{n}(M)T^{D-n+1}-T\left(\zeta(M;0)\ln T+\frac{1}{2}\zeta'(M;0)\right)\\
&-\Bigl( \psi(1)+\ln (4\pi )+\ln( T/\mu)\Bigr)\text{Res}_{s=-\frac{1}{2}}\zeta(M;s)- \sum_{n=D+2}^{\infty}\frac{1}{(2\pi)^{n-D}} \Gamma\left(\frac{n-D}{2}\right)\zeta_R(n-D) c_{n}(M)\frac{1}{T^{n-D-1}}.
\end{split}
\end{equation}
Here $\zeta_R(s)=\sum_{n=1}^{\infty}n^{-s}$ is the Riemann zeta function. On the other hand, it is well known that (see e.g. \cite{15}):
\begin{gather*}
\zeta_{T}(M;0)=-\frac{1}{T}\text{Res}_{s=-\frac{1}{2}}\zeta(M;s),\\
\text{Res}_{s=-\frac{1}{2}} \zeta(M;s)=-\frac{c_{D+1}(M)}{2\sqrt{\pi}},\\
\zeta(M;0)=c_D(M).
\end{gather*}
Hence, the regularized zeta function \eqref{eq10_17_8} is well defined if and only if $c_{D+1}(M)=0$.

In this article, we are going to consider the Casimir free energy of a spherical shell in $(D+1)$-dimensional Minkowski spacetime with radius $r=a$. For regularization purpose, we need to enclose this spherical shell in a larger spherical shell of radius $r=b$. Let $A_{a,b}$ be the annular region $a<r<b$ between the two spherical shells.
The regularized Casimir free energy of the  spherical shell with radius $r=a$ is defined as
\begin{align*}
E_{\text{Cas}}^{\text{reg}}=\lim_{b\rightarrow\infty}\left(E_{\text{Cas}}^{\text{reg}}(B_a)+E_{\text{Cas}}^{\text{reg}}(A_{a,b})-E_{\text{Cas}}^{\text{reg}}(B_b)\right),
\end{align*}where $B_r$ is the ball of radius $r$. Namely, we take the sum of the   energy inside the spherical shell  and the   energy outside the shell, and subtract away the energy when the shell is absent.

Let
\begin{align*}
\zeta_T(a;s)=&\lim_{b\rightarrow\infty}\left(\zeta_T(B_a;s)+\zeta_T(A_{a,b};s)-\zeta_T(B_b;s)\right),\\
\zeta(a;s)=&\lim_{b\rightarrow\infty}\left(\zeta(B_a;s)+\zeta(A_{a,b};s)-\zeta(B_b;s)\right),
\end{align*}
and
\begin{align*}
\hat{c}_n=\lim_{b\rightarrow\infty}\left(c_n(B_a)+c_n(A_{a,b})-c_n(B_b)\right).
\end{align*}
Denote $\zeta(B_r;s)$ by $\zeta^{\text{int}}(r;s)$. As in \cite{10}, one finds that $ \zeta(A_{a,b};s)$ can be written as a sum of three terms:
$$ \zeta(A_{a,b};s)=\zeta^{\text{ext}}(a;s)+\zeta^{\text{inter}}(a,b;s)+\zeta^{\text{int}}(b;s),$$
where the term $\zeta^{\text{ext}}(a;s)$ only depends on $a$, and $\zeta^{\text{inter}}(a,b;s)$ is the interacting term which is regular for all $s$ and goes to zero in the limit $b\rightarrow \infty$. Hence,
\begin{align*}
\zeta(a;s)=\zeta^{\text{int}}(a;s)+\zeta^{\text{ext}}(a;s),
\end{align*}
and
\begin{align*}
\hat{c}_n=\text{Res}_{s=\frac{D-n}{2}}\left(\Gamma(s) \zeta (a;s) \right).
\end{align*}$\zeta^{\text{int}}(a;s)$ can be considered as contribution from the interior region, and $\zeta^{\text{ext}}(a;s)$ can be considered as contribution from the exterior region.

In the same way, one can show that the thermal zeta function can be written as
\begin{align*}
\zeta_T(a;s)=\zeta^{\text{int}}_T(a;s)+\zeta^{\text{ext}}_T(a;s),
\end{align*}and the regularized Casimir free energy is given by
\begin{equation}\label{eq4_23_1}
E_{\text{Cas}}^{\text{reg}}  =-\frac{T}{2}\left(\zeta_{T}'(a;0)+\ln [\mu ]^2\zeta_{T}(a;0)\right).
\end{equation}
As discussed above, the regularized Casimir free energy is free of ambiguities if and only if $\hat{c}_{D+1}=0$. In the high temperature limit, the Casimir free energy has an asymptotic expansion
\begin{equation}\label{eq4_8_1}
\begin{split}
  E_{\text{Cas}}^{\text{reg}} \sim &-\frac{1}{\sqrt{\pi}}\sum_{n=0}^{D-1}2^{D-n} \Gamma\left(\frac{D-n+1}{2}\right)\zeta_R(D-n+1) \hat{c}_{n} T^{D-n+1}-T\left(\zeta(a; 0)\ln T+\frac{1}{2}\zeta'(a; 0)\right)\\
&+\frac{ \psi(1)+\ln (4\pi )+\ln( T/\mu)}{\sqrt{2\pi}}\hat{c}_{D+1}- \sum_{n=D+2}^{\infty}\frac{1}{(2\pi)^{n-D}} \Gamma\left(\frac{n-D}{2}\right)\zeta_R(n-D) \hat{c}_{n} \frac{1}{T^{n-D-1}}.
\end{split}
\end{equation}As was discussed in \cite{9,8}, we need to renormalize this Casimir free energy by subtracting away all the terms of order $T^2$ and above, so that in the high temperature limit, the renormalized Casimir free energy behaves like
\begin{equation}\label{eq5_7_1}
\begin{split}
  E_{\text{Cas}}^{\text{ren}} \sim & -T\left(\zeta(a; 0)\ln T+\frac{1}{2}\zeta'(a; 0)\right)
 +\frac{ \psi(1)+\ln (4\pi )+\ln( T/\mu)}{\sqrt{2\pi}}\hat{c}_{D+1}\\&- \sum_{n=D+2}^{\infty}\frac{1}{(2\pi)^{n-D}} \Gamma\left(\frac{n-D}{2}\right)\zeta_R(n-D) \hat{c}_{n} \frac{1}{T^{n-D-1}}.
\end{split}
\end{equation}
In other words,
\begin{align}\label{eq4_23_2}
E_{\text{Cas}}^{\text{ren}}=&E_{\text{Cas}}^{\text{reg}}+\frac{1}{\sqrt{\pi}}\sum_{n=0}^{D-1}2^{D-n} \Gamma\left(\frac{D-n+1}{2}\right)\zeta_R(D-n+1) \hat{c}_{n} T^{D-n+1}.
\end{align}
From this, we see that the heat kernel coefficients $\hat{c}_{n}$, $0\leq n\leq D-1$, are important for the renormalization of the Casimir free energy. The vanishing of the coefficient $\hat{c}_{D+1}$ will render the Casimir free energy well-defined. After renormalization, in the limit $T\gg 1$,
\begin{equation}\label{eq5_7_2}
\begin{split}
  E_{\text{Cas}}^{\text{ren}} \sim & -T\left(\zeta(a; 0)\ln T+\frac{1}{2}\zeta'(a; 0)\right)
 +\frac{ \psi(1)+\ln (4\pi )+\ln( T/\mu)}{\sqrt{2\pi}}\hat{c}_{D+1}.
\end{split}
\end{equation}The leading term is of order $T\ln T$ with coefficient $-\hat{c}_D$. The next-to-leading order term is of order $T$ with coefficient
$$-\frac{1}{2}\zeta'(a;0).$$ Finally, there are terms of order $\ln T$ and $T^0$ if and only if $\hat{c}_{D+1}=0$, i.e., when the Casimir free energy is not well-defined. In this work, our main goal is to compute the heat kernel coefficients $\hat{c}_n$, $0\leq n\leq D+1$,  the derivative  of the thermal zeta function $\zeta_T(a;s)$ at $s=0$ and the derivative of the zeta function $\zeta(a;s)$ at $s=0$.

The spectrum of a scalar field in a spherical symmetric cavity subject to Dirichlet (D) or Neumann (N) boundary conditions is well known (see e.g. \cite{1}). The spectrum of an electromagnetic field in a spherical symmetric cavity with   perfectly conducting (PC) or infinitely permeable (IP) boundary conditions was studied in \cite{10}. From the spectrum, one can construct the corresponding zeta functions.  For scalar field with Dirichlet   or Neumann   boundary conditions,
\begin{equation}\label{eq4_9_5}
\zeta_{\text{D/N}}(a;s)= \sum_{l=0}^{\infty}b_D(l)\zeta_{\text{D}/ \left(\text{R},\tfrac{2-D}{2}\right) }^{ l+\frac{D-2}{2}}(a;s;0),\end{equation}where
\begin{align*}
b_D(l)= \frac{(2l+D-2)(l+D-3)!}{(D-2)! l!},
\end{align*}
\begin{equation*}
\begin{split}
\zeta_{\text{D}}^{\nu}(a;s;m)=& \frac{\sin (\pi s)}{\pi}\int_m^{\infty}dz(z^2-m^2)^{-s} \frac{d}{dz}\ln\Bigl\{   I_{\nu}(az)K_{\nu}(az) \Bigr\},\\
\zeta_{\text{R},c}^{ \nu}(a;s;m)=& \frac{\sin (\pi s)}{\pi }\int_m^{\infty}dz(z^2-m^2)^{-s}  \frac{d}{dz}\ln\Bigl\{z^{\chi(c,\nu)}  \left[cI_{\nu}(az)+azI_{\nu}'(az)\right] \left[-cK_{\nu}(az)-azK_{\nu}'(az)\right] \Bigr\}.
\end{split}
\end{equation*}Here $\chi(c,\nu)=-2$ when  $\nu+c=0$,   $\chi(c,\nu)=0$ when $\nu+c\neq 0$. The part that involves the modified Bessel function $I_{\nu}(z)$ comes from the interior region, and the part that involves the modified Bessel function $K_{\nu}(z)$ comes from the exterior region.

Notice that $\chi(c,\nu)= 0$ only happens for   Neumann boundary conditions, and it only affects the term with $l=0$ for which $\nu=(D-2)/2$. In this case,
 \begin{align*}
 z^{\chi(c,\nu)}  \left[cI_{\nu}(az)+azI_{\nu}'(az)\right] \left[-cK_{\nu}(az)-azK_{\nu}'(az)\right]=I_{\nu+1}(az)K_{\nu+1}(az).
 \end{align*}Hence,
 \begin{equation}\label{eq4_18_3}
\zeta_{\text{N}}(a;s)= \sum_{l=1}^{\infty}b_D(l)\zeta_{  \left(\text{R},\tfrac{2-D}{2}\right) }^{ l+\frac{D-2}{2}}(a;s;0)
+ \zeta_{\text{D}}^{\tfrac{D}{2}}(a;s;0),\end{equation}where we have used the fact that $b_D(0)=1$.

For electromagnetic field with perfectly conducting   or infinitely permeable boundary conditions, one can divide the spectrum into contributions from transverse electric (TE) modes and from transverse magnetic (TM) modes, and the corresponding zeta function can also be written as a sum of the TE contribution and the TM contribution, namely
\begin{equation}\label{eq4_9_6}\zeta_{\text{PC/IP}} (a;s)=\zeta_{\text{PC/IP, TE}} (a;s)+\zeta_{\text{PC/IP, TM}} (a;s).\end{equation}
 For perfectly conducting boundary conditions,
\begin{equation}\label{eq3_3_8}
\begin{split}
 \zeta_{\text{PC,TE}} (a;s)
=&\sum_{l=1}^{\infty}h_D(l)\zeta_{\text{D}}^{ l+\frac{D-2}{2}}(a;s;0),\\
\zeta_{\text{PC, TM}}  (a;s)=&
 \sum_{l=1}^{\infty}b_D(l)\zeta_{\text{R},\frac{D-2}{2}}^{ l+\frac{D-2}{2}}(a;s;0).
\end{split}
\end{equation} For infinitely permeable boundary conditions,
\begin{equation}\label{eq3_3_9}
\begin{split}
 \zeta_{\text{IP, TE}}  (a;s)=&\sum_{l=1}^{\infty}h_D(l)\zeta_{\text{R},\frac{4-D}{2}}^{ l+\frac{D-2}{2}}(a;s;0),\\
 \zeta_{\text{IP, TM}}  (a;s)=&\sum_{l=1}^{\infty}b_D(l)\zeta_{\text{D}}^{ l+\frac{D-2}{2}}(a;s;0).
\end{split}
\end{equation}Here
$$h_{D}(l)=\frac{l(l+D-2)(2l+D-2)(l+D-4)!}{(D-3)!(l+1)!}. $$

The corresponding thermal zeta functions can be written down in the same way. For example, for scalar field with Dirichlet boundary conditions,
\begin{align*}
\zeta_{T, \text{D}}(a;s)=&\sum_{l=0}^{\infty}\sum_{p=-\infty}^{\infty}b_D(l)\zeta_{   \text{D}  }^{l+\frac{D-2}{2}}(a;s;2\pi pT),\end{align*} Namely, one replace the $m=0$ with $m=2\pi pT$ and sum over $p$ from $-\infty$ to $\infty$. From this, we see that the basic constituents of the zeta functions and the thermal zeta functions are   $ \zeta_{\text{D}}^{\nu}(a;s;m)$ and $ \zeta_{\text{R}, c}^{\nu}(a;s;m)$.

\section{Analytic continuations of zeta functions}

The analytic continuations of the zeta functions $\zeta_{\text{D}} (a;s )$ and $\zeta_{N} (a;s )$ have been discussed in a number of works. See e.g. \cite{1,3,7,12,13,5}. Here we apply the same method to perform the analytic continuations of the zeta functions $\zeta_{\text{PC}} (a;s )$ and $\zeta_{\text{IP} } (a;s )$ and the  thermal zeta functions.

 First consider the functions $\zeta_{\text{D}}^{ \nu}(a;s;m)$ and $\zeta_{\text{R},c}^{  \nu}(a;s;m)$. Debye uniform asymptotic expansions of the modified Bessel function states that \cite{14}: As $\nu\rightarrow \infty$,
\begin{equation}\label{eq3_4_1}\begin{split}
I_{\nu}(\nu z) \sim &\frac{1}{\sqrt{2\pi \nu}}\frac{e^{\nu\eta(z)}}{(1+z^2)^{1/4}} \sum_{k=0}^{\infty} \frac{u_k(t(z))}{\nu^k},\\
I_{\nu}'(\nu z)\sim & \frac{1}{\sqrt{2\pi \nu}}\frac{e^{\nu\eta(z)}(1+z^2)^{1/4}}{z}\sum_{k=0}^{\infty} \frac{v_k(t(z))}{\nu^k},\\
K_{\nu}(\nu z) \sim &\sqrt{\frac{\pi}{2 \nu}}\frac{e^{-\nu\eta(z)}}{(1+z^2)^{1/4}} \sum_{k=0}^{\infty} (-1)^k\frac{u_k(t(z))}{\nu^k},\\
K_{\nu}'(\nu z)\sim & -\sqrt{\frac{\pi}{2 \nu}}\frac{e^{-\nu\eta(z)}(1+z^2)^{1/4}}{z}\sum_{k=0}^{\infty}(-1)^k \frac{v_k(t(z))}{\nu^k},
\end{split}\end{equation}where
\begin{align*}
\eta(z)=\sqrt{1+z^2}+\ln\frac{z}{1+\sqrt{1+z^2}},\hspace{1cm}t(z)=\frac{1}{\sqrt{1+z^2}},
\end{align*}and $u_k(t)$ and $v_k(t)$ are defined recursively by
\begin{align*}
&u_0(t)=1, \hspace{0.5cm}u_{k}(t)=\frac{t^2(1-t^2)}{2}u_{k-1}'(t)+\frac{1}{8}\int_0^td\tau(1-5\tau^2)u_{k-1}(\tau),\\
&v_0(t)=1,\hspace{0.5cm} v_{k }(t)=u_k(t)-t^2(1-t^2)u_{k-1}'(t)-\frac{t(1-t^2)}{2}u_{k-1}(t).
\end{align*}It follows that
\begin{equation*}\begin{split}
cI_{\nu}(\nu z)+\nu zI_{\nu}'(\nu z)\sim & \frac{\sqrt{\nu}e^{\nu\eta(z)}(1+z^2)^{1/4}}{\sqrt{2\pi}}\left(1+\sum_{k=1}^{\infty} \frac{ct(z)u_{k-1}(t(z))+v_k(t(z))}{\nu^k}\right),\\
cK_{\nu}(\nu z)+\nu zK_{\nu}'(\nu z)\sim &- \sqrt{\frac{\pi\nu}{2}}  e^{-\nu\eta(z)}(1+z^2)^{1/4}\left(1+\sum_{k=1}^{\infty}(-1)^k \frac{ct(z)u_{k-1}(t(z))+v_k(t(z))}{\nu^k}\right).\end{split}
\end{equation*}
  Let
\begin{equation}\label{eq3_4_5}
\begin{split}
\ln\left(1+\sum_{k=1}^{\infty} \frac{u_k(t )}{\nu^k}\right)=\sum_{n=1}^{\infty}\frac{D_n(t)}{\nu^n},\hspace{1cm}\ln
\left(1+\sum_{k=1}^{\infty} \frac{ct u_{k-1}(t )+v_k(t )}{\nu^k}\right)=\sum_{n=1}^{\infty} \frac{M_{n,c}(t)}{\nu^n}.
\end{split}
\end{equation}
Then
\begin{equation}\label{eq3_15_3}
\begin{split}
\ln\left(1+\sum_{k=1}^{\infty} (-1)^k\frac{u_k(t )}{\nu^k}\right)=\sum_{n=1}^{\infty}(-1)^n\frac{D_n(t)}{\nu^n},\hspace{1cm}\ln
\left(1+\sum_{k=1}^{\infty}(-1)^k \frac{ct u_{k-1}(t )+v_k(t )}{\nu^k}\right)=\sum_{n=1}^{\infty} (-1)^n\frac{M_{n,c}(t)}{\nu^n}.
\end{split}
\end{equation}$D_n(t)$ and $M_n(t)$ can be computed recursively by
\begin{align*}
D_1(t)=u_1(t),\hspace{1cm} &D_n(t)=u_n(t)-\frac{1}{n}\sum_{j=1}^{n-1}jD_j(t)u_{n-j}(t),\\
M_{1,c}(t)=ct+v_1(t),\hspace{0.5cm}&M_{n,c}(t)=ctu_{n-1}(t)+v_n(t)-\frac{1}{n}\sum_{j=1}^{n-1}jM_{j,c}(t)\left(ctu_{n-j-1}(t)+v_{n-j}(t)\right).
\end{align*}One can   prove by induction that they are polynomials in $t$ of the form
\begin{align*}
D_n(t)=\sum_{k=0}^n d_{n,k}t^{n+2k}, \hspace{1cm}M_n(t)=\sum_{k=0}^n m_{n,k}(c)t^{n+2k}.
\end{align*}
From the Debye asymptotic expansions, we find that
\begin{equation}\label{eq3_4_2}
\begin{split}
&\frac{d}{dz}\ln\Bigl\{ I_{\nu}(\nu z)K_{\nu}(\nu z)\Bigr\} \\=&  -\frac{z}{1+z^2}-2\sum_{i=1}^N \frac{zt(z)^3D_{2i}'(t(z))}{\nu^{2i}}+\frac{d}{dz}\left(\ln \left(I_{\nu}(\nu z)K_{\nu}(z)\right)-\ln\frac{1}{ 2  \nu (1+z^2)^{1/2}} -2\sum_{i=1}^N \frac{D_{2i}(t(z))}{\nu^{2i}}\right),\end{split}\end{equation}
and
\begin{equation}\label{eq4_18_1}
\begin{split}
 &\frac{d}{dz}\ln\Bigl\{  \left[cI_{\nu}(\nu z)+\nu zI_{\nu}'(\nu z)\right]\left[-cK_{\nu}(az)-azK_{\nu}'(az)\right]\Bigr\}
\\\sim &  \frac{z}{1+z^2}-2\sum_{i=1}^N \frac{zt(z)^3M_{2i,c}'(t(z))}{\nu^{2i}}\\&+\frac{d}{dz}\left(\ln \Bigl\{  \left[cI_{\nu}(\nu z)+\nu zI_{\nu}'(\nu z)\right]\left[-cK_{\nu}(az)-azK_{\nu}'(az)\right]\Bigr\}-\ln\frac{ \nu(1+z^2)^{1/2}}{2} -2\sum_{i=1}^N \frac{M_{2i,c}(t(z))}{\nu^{2i}}\right),
\end{split}\end{equation}where $N$ is any positive integer.

 Using this, we can write the zeta functions $\zeta_{\text{D}/ (\text{R},c)}^{\nu} (a;s;m)$  as
\begin{equation}\label{eq4_9_4}
\begin{split}
 \zeta_{\text{D}/ (\text{R},c)}^{\nu} (a;s;m)=\sum_{i=0}^N A_{\text{D}/ (\text{R},c),i}^{ \nu}(a;s;m)+B_{\text{D}/ (\text{R},c),N}^{ \nu}(a;s;m),
\end{split}
\end{equation}where
\begin{equation}\label{eq4_9_1}
\begin{split}
-A_{\text{D},0}^{ \nu}(a;s;m)=&A_{(\text{R},c),0}^{\text{int},\nu}(a;s;m)\\=& \frac{\sin (\pi s)}{\pi}\int_{\frac{a m}{\nu}}^{\infty} dz\left(\left[\frac{\nu z}{a}\right]^2-m^2\right)^{-s} \frac{z}{1+z^2}\\
=& \frac{1}{2}a^{2s}\left(\nu^2+a^2m^2\right)^{-s};
\end{split}
\end{equation}  for $i\geq 1$,
\begin{equation}\label{eq4_9_2}\begin{split}
A_{\text{D},i}^{ \nu}(a;s;m) =&- \frac{2\sin (\pi s)}{\pi}\int_{\frac{a m}{\nu}}^{\infty}dz\left(\left[\frac{\nu z}{a}\right]^2-m^2\right)^{-s}\left\{\frac{zt(z)^3D_{2i}'(t(z))}{\nu^{2i}}\right\}\\
=&-2a^{2s}\sum_{k=0}^{2i}d_{2i,k}\frac{\Gamma\left(s+i+k\right)}{\Gamma(s)\Gamma\left(i+k\right)}\nu^{2k}(\nu^2+a^2m^2)^{-s-i-k},\\
A_{(\text{R},c),i}^{\text{int},\nu}(s) =&- \frac{\sin (\pi s)}{\pi}\int_{\frac{a m}{\nu}}^{\infty}dz\left(\left[\frac{\nu z}{a}\right]^2-m^2\right)^{-s}\left\{\frac{zt(z)^3M_{2i,c}'(t(z))}{\nu^{2i}}\right\}\\
=&-2a^{2s}\sum_{k=0}^{2i}m_{2i,k}(c)\frac{\Gamma\left(s+i+k\right)}{\Gamma(s)\Gamma\left(i+k\right)}\nu^{2k}(\nu^2+a^2m^2)^{-s-i-k};\end{split}
\end{equation}
and
\begin{equation}\label{eq4_9_3}
\begin{split}
B_{\text{D},N}^{ \nu}(a;s;m)=& \frac{\sin (\pi s)}{\pi}\int_{\frac{a m}{\nu}}^{\infty}dz\left(\left[\frac{\nu z}{a}\right]^2-m^2\right)^{-s}
\frac{d}{dz}\left(\ln \left(I_{\nu}(\nu z)K_{\nu}(\nu z)\right)-\ln\frac{1}{ 2  \nu (1+z^2)^{1/2}} -2\sum_{i=1}^N \frac{D_{2i}(t(z))}{\nu^{2i}}\right),\end{split}\end{equation}\begin{equation}\begin{split}
&B_{(\text{R},c),N}^{ \nu}(a;s;m)\\=& \frac{\sin (\pi s)}{\pi}\int_{\frac{a m}{\nu}}^{\infty}dz\left(\left[\frac{\nu z}{a}\right]^2-m^2\right)^{-s}
\\& \times\frac{d}{dz}\left(\ln \left\{  \left[cI_{\nu}(\nu z)+\nu zI_{\nu}'(\nu z)\right]\left[-cK_{\nu}(\nu z)-\nu zK_{\nu}'(\nu z)\right]\right\}-\ln\frac{ \nu(1+z^2)^{1/2}}{2} -2\sum_{i=1}^N \frac{M_{2i,c}(t(z))}{\nu^{2i}}\right).
\end{split}\end{equation}
Now notice that $b_D(l)$ and $h_D(l)$ can be expanded as
\begin{equation*}\begin{split}
h_{D}(l)=& \sum_{j=0}^{ D-2} x_{D;j}\left(l+\frac{D-2}{2}\right)^j,\\
 b_D(l) = &  \sum_{j=1}^{ D-2} y_{D;j}\left(l+\frac{D-2}{2}\right)^j.\end{split}
\end{equation*}
 Let
\begin{equation*}
\zeta_H(s;\chi)=\sum_{n=0}^{\infty}(n+\chi)^{-s}
\end{equation*}be the Hurwitz zeta function.
Substituting \eqref{eq4_9_1}, \eqref{eq4_9_2} and \eqref{eq4_9_3} into \eqref{eq4_9_4}, we find that the zeta functions \eqref{eq4_9_5}, \eqref{eq4_18_3}, \eqref{eq3_3_8} and \eqref{eq3_3_9} can be written as
\begin{equation}\label{eq4_10_1}\begin{split}
\zeta_{\text{D}}(a;s)=&-\frac{a^{2s}}{2}\sum_{j=0}^{D-2}y_{D;j}\zeta_H\left(2s-j;\tfrac{D-2}{2}\right)-2a^{2s}\sum_{j=0}^{D-2}y_{D;j}\sum_{i=1}^N  \zeta_H\left(2s+2i-j;\tfrac{D-2}{2}\right)\sum_{k=0}^{2i} d_{2i,k} \frac{\Gamma\left(s+i+k\right)}{\Gamma(s)\Gamma\left(i+k\right)}+B_{\text{D}}^N(a;s),\\
\zeta_{\text{N}}(a;s)=&\frac{a^{2s}}{2}\sum_{j=0}^{D-2}y_{D;j}\zeta_H\left(2s-j;\tfrac{D}{2}\right)-2a^{2s}\sum_{j=0}^{D-2}y_{D;j}\zeta_H\left(2s+2i-j;\tfrac{D}{2}\right)\sum_{i=1}^N \sum_{k=0}^{2i} m_{2i,k}\left(\tfrac{2-D}{2}\right) \frac{\Gamma\left(s+i+k\right)}{\Gamma(s)\Gamma\left(i+k\right)} +B_{\text{N}}^N(a;s)\\
&-\frac{1}{2}a^{2s}\left(\tfrac{D}{2}\right)^{-2s}+B_{\text{D},0}^{\tfrac{D}{2}}(a;s;0)
,\\
\zeta_{\text{PC}}(a;s) =&  -\frac{a^{2s}}{2}\sum_{j=0}^{D-2}\left(x_{D;j}-y_{D;j}\right)\zeta_H\left( 2s-j;\tfrac{D}{2}\right)\\&-2a^{2s}\sum_{j=0}^{D-2}\sum_{i=1}^N\zeta_H\left(2s+2i-j;\tfrac{D}{2}\right)\sum_{k=0}^{2i}\left(x_{D;j}d_{2i,k}
+y_{D;j}m_{2i,k}\left(\tfrac{D-2}{2}\right)\right)\frac{\Gamma\left(s+i+k\right)}
{\Gamma(s)\Gamma\left(i+k\right)} +B^{N}_{\text{PC}}(a;s),
\\
\zeta_{\text{IP}}(a;s)=& \frac{a^{2s}}{2}\sum_{j=0}^{D-2}\left(x_{D;j}-y_{D;j}\right)\zeta_H\left( 2s-j;\tfrac{D}{2}\right)\\&-2a^{2s}\sum_{j=0}^{D-2}\sum_{i=1}^N\zeta_H\left(2s+2i-j;\tfrac{D}{2}\right)\sum_{k=0}^{2i}\left(x_{D;j}m_{2i,k}
\left(\tfrac{4-D}{2}\right)+y_{D;j}d_{2i,k} \right)\frac{\Gamma\left(s+i+k\right)}
{\Gamma(s)\Gamma\left(i+k\right)}    +B^{N}_{\text{IP}}(a;s).
\end{split}
\end{equation}Here
\begin{equation*}\begin{split}
B^{N}_{\text{D}}(a;s)=& \sum_{l=0}^{\infty}b_D(l)B_{\text{D},N}^{ l+\frac{D-2}{2}}(a;s;0),\\
B^{N}_{\text{N}}(a;s)=& \sum_{l=1}^{\infty}b_D(l)B_{\left(\text{R},\tfrac{2-D}{2}\right),N}^{ l+\frac{D-2}{2}}(a;s;0),\\
B^{N}_{\text{PC}}(a;s)=& \sum_{l=1}^{\infty}h_D(l) B_{\text{D},N}^{ l+\frac{D-2}{2}}(a;s;0)+ \sum_{l=1}^{\infty}b_D(l) B_{\left(\text{R},\tfrac{D-2}{2}\right),N}^{ l+\frac{D-2}{2}}(a;s;0),\\
B^{N}_{\text{PC}}(a;s)=&\sum_{l=1}^{\infty}h_D(l) B_{\left(\text{R},\tfrac{4-D}{2}\right),N}^{ l+\frac{D-2}{2}}(a;s;0)+ \sum_{l=1}^{\infty}b_D(l) B_{\text{D},N}^{ l+\frac{D-2}{2}}(a;s;0).
\end{split}\end{equation*} By taking $N\geq D/2$, one can guarantee that $B^{N}_{\text{D}/N/\text{PC}/\text{IP}}(a;s)$ does not contain any pole  on the half-plane $\text{Re}\;s>-1$. Hence, this gives the analytic continuation of the zeta functions to $\text{Re}\;s>-1$.

Next, define
\begin{equation}\label{eq4_22_1}
\Xi(s,\alpha;\chi;c)=\sum_{n=0}^{\infty}\sum_{p=-\infty}^{\infty}\frac{(n+\chi)^{\alpha}}{\left((n+\chi)^2+(cp)^2\right)^{s}},
\end{equation}
\begin{align}\label{eq4_22_2}
X(s)=\sum_{p=-\infty}^{\infty}\frac{1}{\left(\left[\tfrac{D}{2}\right]^2+[2\pi p aT]^2\right)^{s}}.
\end{align}We find in the same way that the thermal zeta functions can be written as
\begin{equation}\label{eq4_22_3}\begin{split}
\zeta_{T,\text{D}}(a;s)=&-\frac{a^{2s}}{2}\sum_{j=0}^{D-2}y_{D;j}\Xi\left(s,j;\tfrac{D-2}{2};2\pi aT\right)\\&-2a^{2s}\sum_{j=0}^{D-2}y_{D;j}\sum_{i=1}^N  \sum_{k=0}^{2i} d_{2i,k} \frac{\Gamma\left(s+i+k\right)}{\Gamma(s)\Gamma\left(i+k\right)}\Xi\left(s+i+k,2k+j;\tfrac{D-2}{2};2\pi aT\right)+E_{\text{D}}^N(a;s),\\
\zeta_{T,\text{N}}(a;s)=&\frac{a^{2s}}{2}\sum_{j=0}^{D-2}y_{D;j}\Xi\left(s,j;\tfrac{D}{2};2\pi aT\right)\\&-2a^{2s}\sum_{j=0}^{D-2}y_{D;j} \sum_{i=1}^N \sum_{k=0}^{2i} m_{2i,k}\left(\tfrac{2-D}{2}\right) \frac{\Gamma\left(s+i+k\right)}{\Gamma(s)\Gamma\left(i+k\right)}\Xi\left(s+i+k,2k+j;\tfrac{D}{2};2\pi aT\right) +E_{\text{N}}^N(a;s)\\
&-\frac{a^{2s}}{2}X(s)-2a^{2s}\sum_{k=0}^2d_{2,k}
\frac{\Gamma\left(s+1+k\right)}{\Gamma(s)\Gamma\left(1+k\right)} \left(\tfrac{D}{2}\right)^{2k}X(s+k+1)+\sum_{p=-\infty}^{\infty}B_{\text{D},1}^{ \frac{D}{2}}(a;s;2\pi p T),
\\
\zeta_{T,\text{PC}}(a;s) =&  -\frac{a^{2s}}{2}\sum_{j=0}^{D-2}\left(x_{D;j}-y_{D;j}\right)\Xi\left(s,j;\tfrac{D}{2};2\pi aT\right)\\&-2a^{2s}\sum_{j=0}^{D-2}\sum_{i=1}^N \sum_{k=0}^{2i}\left(x_{D;j}d_{2i,k}+y_{D;j}m_{2i,k}\left(\tfrac{D-2}{2}\right)\right)\frac{\Gamma\left(s+i+k\right)}
{\Gamma(s)\Gamma\left(i+k\right)}\Xi\left(s+i+k,2k+j;\tfrac{D}{2};2\pi aT\right)   +E^{N}_{\text{PC}}(a;s),
\\
\zeta_{T,\text{IP}}(a;s)=& \frac{a^{2s}}{2}\sum_{j=0}^{D-2}\left(x_{D;j}-y_{D;j}\right)\Xi\left(s,j;\tfrac{D}{2};2\pi aT\right)\\&-2a^{2s}\sum_{j=0}^{D-2}\sum_{i=1}^N \sum_{k=0}^{2i}\left(x_{D;j}m_{2i,k}\left(\tfrac{4-D}{2}\right)+y_{D;j}d_{2i,k} \right)
\frac{\Gamma\left(s+i+k\right)}
{\Gamma(s)\Gamma\left(i+k\right)}\Xi\left(s+i+k,2k+j;\tfrac{D}{2};2\pi aT\right)     +E^{N}_{\text{IP}}(a;s).
\end{split}
\end{equation}Here
\begin{equation*}\begin{split}
E^{N}_{\text{D/N}}(a;s)=& \sum_{l=0}^{\infty}\sum_{p=-\infty}^{\infty}b_D(l)B_{\text{D},N}^{ l+\frac{D-2}{2}}(a;s;2\pi p T),\\
E^{N}_{\text{N}}(a;s)=& \sum_{l=1}^{\infty}\sum_{p=-\infty}^{\infty}b_D(l)B_{\left(\text{R},\tfrac{2-D}{2}\right),N}^{ l+\frac{D-2}{2}}(a;s;2\pi p T),\\
E^{N}_{\text{PC}}(a;s)=& \sum_{l=1}^{\infty}\sum_{p=-\infty}^{\infty} h_D(l) B_{\text{D},N}^{ l+\frac{D-2}{2}}(a;s;2\pi p T)+ \sum_{l=1}^{\infty}b_D(l) B_{\left(\text{R},\tfrac{D-2}{2}\right),N}^{ l+\frac{D-2}{2}}(a;s;2\pi p T),\\
E^{N}_{\text{PC}}(a;s)=&\sum_{l=1}^{\infty}\sum_{p=-\infty}^{\infty}h_D(l) B_{\left(\text{R},\tfrac{4-D}{2}\right),N}^{ l+\frac{D-2}{2}}(a;s;2\pi p T)+ \sum_{l=1}^{\infty}b_D(l) B_{\text{D},N}^{ l+\frac{D-2}{2}}(a;s;2\pi p T).
\end{split}\end{equation*} By taking $N\geq D/2$, one can guarantee that $E^{N}_{\text{D}/N/\text{PC}/\text{IP}}(a;s)$ does not contain any pole  on the half-plane $\text{Re}\;s>-1/2$. Hence, this gives the analytic continuation of the thermal zeta functions to $\text{Re}\;s>-1/2$.

\section{Heat kernel coefficients}\label{hkc}
Using the fact that $\zeta_H(s;\chi)$ has only one pole at $s=1$ with residue $1$, we read immediately from \eqref{eq4_10_1}  that, for Dirichlet boundary conditions,
\begin{equation*}
\begin{split}
\text{if}\; \;0\leq n\leq & D-1,\\
\hat{c}_n=&a^{D-n}\left(-\frac{y_{D;D-n-1}}{4}\Gamma\left(\tfrac{D-n}{2}\right)-\sum_{i=1}^{\left[\frac{n-1}{2}\right]}y_{D;D-n+2i-1}\sum_{k=0}^{2i} d_{2i,k} \frac{\Gamma\left(\tfrac{D-n}{2}+i+k\right)}{ \Gamma\left(i+k\right)}\right),\\
\hat{c}_{D}=&-\frac{1}{2}\sum_{j=0}^{D-2}y_{D;j}\zeta_H\left(-j;\tfrac{D-2}{2}\right)-\sum_{i=1}^{\left[\frac{D-1}{2}\right]}y_{D;2i-1}\sum_{k=0}^{2i} d_{2i,k},\\
\hat{c}_{D+1}=&-\frac{1}{a}\sum_{i=1}^{\left[\tfrac{D}{2}\right]}y_{D;2i-2}\sum_{k=0}^{2i}d_{2i,k}\frac{\Gamma\left(i+k-\tfrac{1}{2}\right)}{\Gamma(i+k)};
\end{split}
\end{equation*}
\begin{table} \caption{\label{t1}Heat kernel coefficients for Dirichlet boundary condition}

\begin{tabular}{c|cccccc }
\hline
\hline
& $D=3$ & $D=4$ & $D=5$ & $D=6$ &  $D=7$ & $D=8$   \\
\hline
&&&&&&\\
$\hat{c}_1 $ & $\displaystyle-\frac{a^2}{2}$ & $\displaystyle -\frac{\sqrt{\pi}a^3}{8}$ &$\displaystyle -\frac{a^4}{12}$ &$\displaystyle -\frac{ \sqrt{\pi}a^5}{64}$ &$\displaystyle -\frac{a^6}{120} $ &$\displaystyle -\frac{\sqrt{\pi}a^7}{768}$  \\
&&&&&&\\
$\hat{c}_2 $ & $\displaystyle 0$ & $\displaystyle  0$ &$\displaystyle 0$ &$\displaystyle 0$ &$\displaystyle  0$ &$\displaystyle 0$  \\
&&&&&&\\
$\hat{c}_3$ &$\displaystyle -\frac{1}{24}$&$\displaystyle -\frac{11\sqrt{\pi} a}{256}  $&$\displaystyle -\frac{a^2}{16}$&$\displaystyle -\frac{125\sqrt{\pi}a^3}{6144}$&$\displaystyle -\frac{ a^4}{60}$&$\displaystyle -\frac{91\sqrt{\pi}a^5}{24576} $     \\
&&&&&&\\
$\hat{c}_4 $ & $\displaystyle 0$ & $\displaystyle  0$ &$\displaystyle 0$ &$\displaystyle 0$ &$\displaystyle  0$ &$\displaystyle 0$  \\
&&&&&& \\
$\hat{c}_5$ & &$\displaystyle  -\frac{35\sqrt{\pi}}{32768a} $&$\displaystyle \frac{17}{5760} $&$\displaystyle \frac{2159\sqrt{\pi}a}{786432} $&$\displaystyle \frac{a^2}{256} $&$\displaystyle \frac{60179\sqrt{\pi}a^3}{47185920} $ \\
&&&&&&\\
$\hat{c}_6 $ & &  &$\displaystyle 0$ &$\displaystyle 0$ &$\displaystyle  0$ &$\displaystyle 0$  \\
&&&&\\
$\hat{c}_7$ &&&& $\displaystyle \frac{1685 \sqrt{\pi}}{12582912 a} $&$\displaystyle -\frac{367}{967680}$&$\displaystyle -\frac{260699\sqrt{\pi}a}{754974720} $ \\
&&&&&&\\
$\hat{c}_8 $ &  &  & & &$\displaystyle  0$ &$\displaystyle 0$  \\
&&&&\\
$\hat{c}_9$ &&&&&&$\displaystyle- \frac{1059678257581\sqrt{\pi}}{50665495807918080a}$ \\
&&&&\\
\hline
\hline
\end{tabular}

\end{table}

\noindent
for Neumann boundary conditions,
\begin{equation*}
\begin{split}
\text{if}\; \;0\leq n\leq & D-1,\\
\hat{c}_n=&a^{D-n}\left(\frac{y_{D;D-n-1}}{4}\Gamma\left(\tfrac{D-n}{2}\right)-\sum_{i=1}^{\left[\frac{n-1}{2}\right]}y_{D;D-n+2i-1}\sum_{k=0}^{2i} m_{2i,k} \left(\tfrac{2-D}{2}\right) \frac{\Gamma\left(\tfrac{D-n}{2}+i+k\right)}{ \Gamma\left(i+k\right)}\right),\\
\hat{c}_{D}=&\frac{1}{2}\sum_{j=0}^{D-2}y_{D;j}\zeta_H\left(-j;\tfrac{D}{2}\right)-\sum_{i=1}^{\left[\frac{D-1}{2}\right]}y_{D;2i-1}\sum_{k=0}^{2i} m_{2i,k}\left(\tfrac{2-D}{2}\right)-\frac{1}{2},\\
\hat{c}_{D+1}=&-\frac{1}{a}\sum_{i=1}^{\left[\tfrac{D}{2}\right]}y_{D;2i-2}\sum_{k=0}^{2i}m_{2i,k}\left(\tfrac{2-D}{2}\right)\frac{\Gamma\left(i+k-\tfrac{1}{2}\right)}{\Gamma(i+k)};
\end{split}
\end{equation*}
\begin{table} \caption{\label{t2}Heat kernel coefficients for Neumann boundary condition}

\begin{tabular}{c|cccccc }
\hline
\hline
& $D=3$ & $D=4$ & $D=5$ & $D=6$ &  $D=7$ & $D=8$   \\
\hline
&&&&&&\\
$\hat{c}_1 $ & $\displaystyle \frac{a^2}{2}$ & $\displaystyle  \frac{\sqrt{\pi}a^3}{8}$ &$\displaystyle  \frac{a^4}{12}$ &$\displaystyle  \frac{ \sqrt{\pi}a^5}{64}$ &$\displaystyle  \frac{a^6}{120} $ &$\displaystyle \frac{\sqrt{\pi}a^7}{768}$  \\
&&&&&&\\
$\hat{c}_2 $ & $\displaystyle 0$ & $\displaystyle  0$ &$\displaystyle 0$ &$\displaystyle 0$ &$\displaystyle  0$ &$\displaystyle 0$  \\
&&&&&&\\
$\hat{c}_3$ &$\displaystyle  -\frac{17}{24}$&$\displaystyle  \frac{41\sqrt{\pi} a}{256}  $&$\displaystyle  \frac{3a^2}{16}$&$\displaystyle \frac{335\sqrt{\pi}a^3}{6144}$&$\displaystyle  \frac{ a^4}{24}$&$\displaystyle \frac{217\sqrt{\pi}a^5}{24576} $     \\
&&&&&& \\
$\hat{c}_4 $ & $\displaystyle 0$ & $\displaystyle  -1$ &$\displaystyle 0$ &$\displaystyle 0$ &$\displaystyle  0$ &$\displaystyle 0$  \\
&&&&&&\\
$\hat{c}_5$ & &$\displaystyle   \frac{5861\sqrt{\pi}}{32768a} $&$\displaystyle -\frac{3887}{5760} $&$\displaystyle \frac{108007\sqrt{\pi}a}{786432} $&$\displaystyle \frac{37a^2}{256} $&$\displaystyle \frac{1909579\sqrt{\pi}a^3}{47185920} $ \\
&&&&&& \\
$\hat{c}_6 $ & $\displaystyle $ & $\displaystyle  $ &$\displaystyle 0$ &$\displaystyle -1$ &$\displaystyle  0$ &$\displaystyle 0$  \\
&&&&\\
$\hat{c}_7$ &&&& $\displaystyle \frac{1723783 \sqrt{\pi}}{8388608 a} $&$\displaystyle -\frac{676463}{967680}$&$\displaystyle \frac{170051269\sqrt{\pi}a}{1509949440} $ \\
&&&&&& \\
$\hat{c}_8$ & $\displaystyle $ & $\displaystyle  $ &$\displaystyle $ &$\displaystyle $ &$\displaystyle  0$ &$\displaystyle -1$  \\
&&&&\\
$\hat{c}_9$ &&&&&&$\displaystyle \frac{171400283233\sqrt{\pi}}{773094113280a}$ \\
&&&&\\
\hline
\hline
\end{tabular}

\end{table}

\noindent
for perfectly conducting boundary conditions,
\begin{equation*}\begin{split}
\text{if}\; \;0\leq i\leq & D-1,\\
\hat{c}_n=&
a^{D-n}\left(-\frac{x_{D;D-n-1}-y_{D;D-n-1}}{4}
\Gamma\left(\tfrac{D-n}{2}\right)
\right.\\&\left. - \sum_{i=1}^{\left[\frac{n-1}{2}\right]}  \sum_{k=0}^{2i}\Bigl(x_{D; D-n+2i-1} d_{2i,k}+y_{D; D-n+2i-1}m_{2i,k}\left(\tfrac{D-2}{2}\right) \Bigr)\frac{\Gamma\left(\frac{D-n }{2}+i+k\right)}{\Gamma\left(i+k\right)}\right)
,\\
\hat{c}_{D} =&
-\frac{1}{2}\sum_{j=0}^{D-2}\left(x_{D;j}-y_{D;j}\right)\zeta_H\left(-j;\tfrac{D}{2}\right)
 - \sum_{i=1}^{\left[\tfrac{D-1}{2}\right]} \sum_{k=0}^{2i} \Bigl(x_{D;2i-1}d_{2i,k}+y_{D;2i-1}m_{2i,k}\left(\tfrac{D-2}{2}\right)\Bigr),\\
\hat{c}_{D+1} =& -\frac{1}{ a}   \sum_{i=1}^{\left[\tfrac{D}{2}\right]}
\sum_{k=0}^{2i} \Bigl(x_{D;2i-2}d_{2i,k}+y_{D;2i-2}m_{2i,k}\left(\tfrac{D-2}{2}\right)\Bigr)\frac{\Gamma\left(i+k-\frac{1}{2}\right)}{\Gamma\left(i+k\right)};
\end{split}
\end{equation*}
\begin{table} \caption{\label{t3}Heat kernel coefficients for perfectly conducting boundary condition}

\begin{tabular}{c|cccccc }
\hline
\hline
& $D=3$ & $D=4$ & $D=5$ & $D=6$ &  $D=7$ & $D=8$   \\
\hline
&&&&&&\\
$\hat{c}_1 $ & $0$ & $\displaystyle -\frac{\sqrt{\pi}a^3}{8}$ &$\displaystyle -\frac{a^4}{6}$ &$\displaystyle -\frac{3\sqrt{\pi}a^5}{64}$ &$\displaystyle -\frac{a^6}{30} $ &$\displaystyle -\frac{5\sqrt{\pi}a^7}{768}$  \\
&&&&&&\\
$\hat{c}_2 $ & $\displaystyle 0$ & $\displaystyle  0$ &$\displaystyle 0$ &$\displaystyle 0$ &$\displaystyle  0$ &$\displaystyle 0$  \\
&&&&&&\\
$\hat{c}_3$ &$\displaystyle \frac{1}{4}$&$\displaystyle \frac{211\sqrt{\pi} a}{256}  $&$\displaystyle  a^2 $&$\displaystyle \frac{585\sqrt{\pi}a^3}{2048}$&$\displaystyle \frac{5a^4}{24}$&$\displaystyle \frac{1015\sqrt{\pi}a^5}{24576} $     \\
&&&&&& \\
$\hat{c}_4$ &$\displaystyle 0$&$\displaystyle -1$&$\displaystyle 0$&$\displaystyle 0$&$\displaystyle 0$  &$\displaystyle0$\\
&&&&&& \\
$\hat{c}_5$ & &$\displaystyle  \frac{1631\sqrt{\pi}}{32768a} $&$\displaystyle -\frac{899}{1440} $&$\displaystyle \frac{75361\sqrt{\pi}a}{262144} $&$\displaystyle \frac{163a^2}{384} $&$\displaystyle \frac{6994813\sqrt{\pi}a^3}{47185920} $ \\
&&&&&&\\
$\hat{c}_6$ & &&$\displaystyle 0 $&$\displaystyle -1$&$\displaystyle  0$&$\displaystyle0 $   \\
&&&&\\
$\hat{c}_7$ &&&& $\displaystyle \frac{1052991\sqrt{\pi}}{8388608 a} $&$\displaystyle -\frac{340577}{483840}$&$\displaystyle \frac{231850177\sqrt{\pi}a}{1509949440} $ \\
&&&&\\
$\hat{c}_8$ &&&&&$\displaystyle0$&$\displaystyle -1$ \\
&&&&\\
$\hat{c}_9$ &&&&&&$\displaystyle \frac{800416822715749\sqrt{\pi}}{5066549580791808 a}$ \\
&&&&\\
\hline
\hline
\end{tabular}

\end{table}

\noindent
and finally for infinitely permeable boundary condition,
\begin{equation*}\begin{split}
\text{if}\; \;0\leq i\leq & D-1,\\
\hat{c}_n=&
a^{D-n}\left(  \frac{x_{D;D-n-1}-y_{D;D-n-1}}{4}
\Gamma\left(\tfrac{D-n}{2}\right)
\right.\\&\left.-\sum_{i=1}^{\left[\tfrac{n-1}{2}\right]}  \sum_{k=0}^{2i}\Bigl(x_{D; D-n+2i-1} m_{2i,k}\left(\tfrac{4-D}{2}\right)+y_{D; D-n+2i-1}d_{2i,k}\Bigr)\frac{\Gamma\left(\frac{D-n}{2}+i+k\right)}{\Gamma\left(i+k\right)}\right)
,\\
\hat{c}_{D} =&
 \frac{1}{2}\sum_{j=0}^{D-2}\Bigl(x_{D;j}-y_{D;j}\Bigr)\zeta_H\left(-j;\tfrac{D}{2}\right)
 - \sum_{i=1}^{\left[\tfrac{D-1}{2}\right]} \sum_{k=0}^{2i} \left(x_{D;2i-1}m_{2i,k}\left(\tfrac{4-D}{2}\right)+y_{D;2i-1}d_{2i,k} \right),\\
\hat{c}_{D+1} =&   -\frac{1}{a}\sum_{i=1}^{\left[\tfrac{D}{2}\right]}
\sum_{k=0}^{2i} \Bigl(x_{D;2i-2}m_{2i,k}\left(\tfrac{4-D}{2}\right)+y_{D;2i-2}d_{2i,k} \Bigr)\frac{\Gamma\left(i+k-\frac{1}{2}\right)}{\Gamma\left(i+k\right)}.
\end{split}
\end{equation*}

\begin{table} \caption{\label{t4}Heat kernel coefficients for infinitely permeable boundary condition}
\begin{tabular}{c|cccccc }
\hline
\hline
& $D=3$ & $D=4$ & $D=5$ & $D=6$ &  $D=7$ & $D=8$   \\
\hline
&&&&&&\\
$\hat{c}_1 $ & $0$ & $\displaystyle  \frac{\sqrt{\pi}a^3}{8}$ &$\displaystyle  \frac{a^4}{6}$ &$\displaystyle \frac{3\sqrt{\pi}a^5}{64}$ &$\displaystyle \frac{a^6}{30} $ &$\displaystyle \frac{5\sqrt{\pi}a^7}{768}$  \\
&&&&&&\\
$\hat{c}_2 $ & $\displaystyle 0$ & $\displaystyle  0$ &$\displaystyle 0$ &$\displaystyle 0$ &$\displaystyle  0$ &$\displaystyle 0$  \\
&&&&&&\\
$\hat{c}_3$ &$\displaystyle \frac{1}{4}$&$\displaystyle -\frac{121\sqrt{\pi} a}{256}  $&$\displaystyle  -\frac{a^2}{2} $&$\displaystyle -\frac{235\sqrt{\pi}a^3}{2048}$&$\displaystyle -\frac{7a^4}{120}$&$\displaystyle -\frac{133\sqrt{\pi}a^5}{24576} $     \\
&&&&&& \\
$\hat{c}_4$ &$\displaystyle 0$&$\displaystyle 1$&$\displaystyle 0$&$\displaystyle 0$&$\displaystyle 0$  &$\displaystyle0$\\
&&&&&& \\
$\hat{c}_5$ & &$\displaystyle  -\frac{2713\sqrt{\pi}}{32768a} $&$\displaystyle  \frac{989}{1440} $&$\displaystyle -\frac{44071\sqrt{\pi}a}{262144} $&$\displaystyle -\frac{23a^2}{128} $&$\displaystyle -\frac{2036587\sqrt{\pi}a^3}{47185920} $ \\
&&&&&&\\
$\hat{c}_6$ & &&$\displaystyle 0 $&$\displaystyle 1$&$\displaystyle  0$&$\displaystyle0 $   \\
&&&&\\
$\hat{c}_7$ &&&& $\displaystyle -\frac{871339\sqrt{\pi}}{4194304 a} $&$\displaystyle \frac{301727}{483840}$&$\displaystyle -\frac{28775291\sqrt{\pi}a}{188743680} $ \\
&&&&\\
$\hat{c}_8$ &&&&&$\displaystyle0$&$\displaystyle 1$ \\
&&&&\\
$\hat{c}_9$ &&&&&&$\displaystyle -\frac{2500126116950921\sqrt{\pi}}{10133099161583616 a}$ \\
&&&&\\
\hline
\hline
\end{tabular}

\end{table}

Using the fact that for any nonnegative integer $j$, $\displaystyle \zeta_H(-j;\chi)=-\frac{B_{j+1}(\chi)}{j+1},$ where $B_n(x)$ is the Bernoulli polynomial of degree $n$, one can readily compute all the coefficients $\hat{c}_n$ for $0\leq n\leq D+1$. In fact, using the fact that  $x_{D;j}$ and $y_{D;j}$ are nonzero if and only if $D$ and $j$ have the same parity,  one can show that for $0\leq n\leq D-1$, $\hat{c}_n=0$ if $n$ is even and $\hat{c}_{D+1}=0$ if $D$ is odd. The latter implies that the Casimir free energy is well defined when the dimension $D$ is odd.

 The coefficients $\hat{c}_n$ for $3\leq D\leq 8, 1\leq n\leq D+1$, are listed in Tables \ref{t1}, \ref{t2}, \ref{t3} and \ref{t4}. An interesting phenomenon to observe is that for $D=4,6,8$, $\hat{c}_D=0, -1 , -1$ and $1$ respectively for Dirichlet, Neumann, perfectly conducting and infinitely permeable boundary conditions.
We conjecture that this is true for all even $D$.

\section{$\boldsymbol{\zeta}'(a;0)$}\label{zetaprime}

In this section, we compute $\zeta'(a;0)$. First notice that when $N$ is large enough,
\begin{equation}\label{eq4_18_2}
\begin{split}
B_{\text{D},N}^{ \nu\prime}(a;0;m)=&  \int_{\frac{a m}{\nu}}^{\infty}dz
\frac{d}{dz}\left(\ln \left(I_{\nu}(\nu z)K_{\nu}(\nu z)\right)-\ln\frac{1}{ 2  \nu (1+z^2)^{1/2}} -2\sum_{i=1}^N \frac{D_{2i}(t(z))}{\nu^{2i}}\right)\\
=&-\ln \left(I_{\nu}(am)K_{\nu}(am)\right)+\ln\frac{1}{ 2    (\nu^2+a^2m^2)^{1/2}} +2\sum_{i=1}^N \frac{D_{2i}(t(am/\nu))}{\nu^{2i}}.\end{split}\end{equation}
Similarly,\begin{equation}\begin{split}
B_{(\text{R},c),N}^{ \nu\prime}(a;0;m)=&  -\ln \Bigl\{\left[cI_{\nu}(am)+amI_{\nu}'(am)\right]\left[-cK_{\nu}(am)-amK_{\nu}'(am)\right]\Bigr\}\\&+\ln\frac{  (\nu^2+a^2m^2)^{1/2}}{2} +2\sum_{i=1}^N \frac{M_{2i,c}(t(am/\nu))}{\nu^{2i}}.
\end{split}\end{equation}
As $z\rightarrow 0$,
\begin{align*}
I_{\nu}(z)\sim & \frac{1}{\Gamma(\nu+1)}\left(\frac{z}{2}\right)^{\nu},\\
K_{\nu}(z)\sim &\frac{\Gamma(\nu)}{2}\left(\frac{z}{2}\right)^{-\nu}.
\end{align*}
It follows that
\begin{align*}
cI_{\nu}(z)+  zI_{\nu}'(z)\sim & \frac{\nu+c}{\Gamma(\nu+1)}\left(\frac{z}{2}\right)^{\nu},\\
cK_{\nu}(z)+  zK_{\nu}'(z)\sim & -\frac{(\nu-c)\Gamma(\nu)}{2}\left(\frac{z}{2}\right)^{-\nu}.
\end{align*}
Hence,
\begin{align*}
&\lim_{m\rightarrow 0}\Bigl(I_{\nu}(am)K_{\nu}(am)\Bigr)=\frac{1}{2\nu},\\
&\lim_{m\rightarrow 0}\Bigl(\left[cI_{\nu}(am)+amI_{\nu}'(am)\right]\left[-cK_{\nu}(am)-amK_{\nu}'(am)\right] \Bigr)=\frac{\nu^2-c^2}{2\nu}.
\end{align*}

On the other hand, $t(0)=1.$
Therefore,
\begin{align*}
B_{\text{D},N}^{ \nu\prime}(a;0;0)=& 2\sum_{i=1}^N\nu^{-2i}\sum_{k=0}^{2i}d_{2i,k},\\
B_{(\text{R},c),N}^{ \nu\prime}(a;0;0)=& \ln\frac{\nu^2}{\nu^2-c^2}+2\sum_{i=1}^N\nu^{-2i}\sum_{k=0}^{2i}m_{2i,k}(c).
\end{align*}
However,
\begin{align*}
\sum_{k=0}^{2i}d_{2i,k}=0,\quad \sum_{k=0}^{2i}m_{2i,k}(c)=-\frac{c^{2i}}{2i}.
\end{align*}
It follows immediately that $$B^{N\prime}_{\text{D}}(a;0)=0.$$

\begin{table}\caption{\label{t5}$\zeta_{\text{D}}'(a;0)-2\zeta_{\text{D}}(a;0)\ln a$ }
\begin{tabular}{c|c|c }
\hline
\hline
&&\\
\hspace{0.5cm}$D$\hspace{0.5cm} & \hspace{0.5cm}$\zeta_{\text{D}}'(a;0)-2\zeta_{\text{D}}(a;0)\ln a$\hspace{0.5cm} &\hspace{0.5cm} numerical value \hspace{0.5cm}  \\
&&\\
\hline
&&\\
3&   $\displaystyle \zeta_R'(-1)+\frac{1}{12}\ln 2-\frac{3}{16}$ & $\displaystyle   -2.9516\times 10^{-1}$ \\
&&\\
4 &$\displaystyle -\zeta_R'(-2)  $&  $\displaystyle  3.0448\times 10^{-2}$ \\
&& \\
5&$\displaystyle -\frac{1}{24}\zeta_R'(-1)+\frac{7}{24}\zeta_R'(-3)-\frac{11}{2880}\ln 2+\frac{47}{4608}$  &$\displaystyle 1.6014\times 10^{-2}$\\
&& \\
6 & $\displaystyle   \frac{1}{12}\zeta_R'(-2)-\frac{1}{12}\zeta_R'(-4) $&$\displaystyle     -3.2027\times 10^{-3} $ \\
&&\\
7 & $\displaystyle  \frac{3}{640}\zeta_R'(-1)-\frac{7}{192}\zeta_R'(-3)+\frac{31}{1920}\zeta_R'(-5)+\frac{211}{483840}\ln 2-\frac{2641}{2150400 }$&$\displaystyle   -1.9066\times 10^{-3}$   \\
&&\\
8 & $\displaystyle  -\frac{1}{90}\zeta_R'(-2)+\frac{1}{72}\zeta_R'(-4)-\frac{1}{360}\zeta_R'(-6) $&$\displaystyle  4.6559\times 10^{-4}$ \\
&&\\
\hline
\hline
\end{tabular}

\end{table}

\begin{table}\caption{\label{t6}$\zeta_{\text{N}}'(a;0)-2\zeta_{\text{N}}(a;0)\ln a$ }
\begin{tabular}{c|c|c }
\hline
\hline
&&\\
\hspace{0.5cm}$D$\hspace{0.5cm} & \hspace{0.5cm}$\zeta_{\text{N}}'(a;0)-2\zeta_{\text{N}}(a;0)\ln a$\hspace{0.5cm} &\hspace{0.5cm} numerical value \hspace{0.5cm}  \\
&&\\
\hline
&&\\
3&   $\displaystyle -\zeta_R'(-1)-\frac{25}{12}\ln 2+\ln 3 -\frac{7}{16}+2\int_0^{1/2}du\left(\ln\Gamma\left(\tfrac{3}{2}+u\right)
-\ln\Gamma\left(\tfrac{3}{2}-u\right)\right)$ & $\displaystyle -6.0823\times 10^{-1} $ \\
&&\\
4 &$\displaystyle \zeta_R'(-2)+2\ln 2-\frac{3}{2}-2\int_0^{1}du\,u\left(\ln\Gamma\left(2+u\right)
+\ln\Gamma\left(2-u\right)\right)  $&  $\displaystyle -4.8203\times 10^{-1}$ \\
&& \\
5&$\displaystyle \frac{1}{24}\zeta_R'(-1)-\frac{7}{24}\zeta_R'(-3)-\frac{8629}{2880}\ln 2 +\ln  5 -\frac{27901}{23040}$  &$\displaystyle -1.9959\times 10^{-1}$\\
& $\displaystyle +\int_0^{3/2}du\left(u^2-\tfrac{1}{12}\right)\left(\ln\Gamma\left(\tfrac{5}{2}+u\right)
-\ln\Gamma\left(\tfrac{5}{2}-u\right)\right)$&\\
&& \\
6 & $\displaystyle   -\frac{1}{12}\zeta_R'(-2)+\frac{1}{12}\zeta_R'(-4) +\ln 144-\frac{5}{2}
 -\int_0^{2}du\left(\tfrac{u^3}{3}-\tfrac{u}{6}\right)\left(\ln\Gamma\left(3+u\right)+\ln\Gamma\left(3-u\right)\right)$&$\displaystyle   -1.6471\times 10^{-1}$ \\
&&\\
7 & $\displaystyle  -\frac{3}{640}\zeta_R'(-1)+\frac{7}{192}\zeta_R'(-3)-\frac{31}{1920}\zeta_R'(-5)-\frac{ 2419411}{483840}\ln 2
+\ln 7-\ln 3 -\frac{ 545849}{307200 }$&$\displaystyle -2.2863 \times 10^{-2}$   \\
& $\displaystyle +\int_0^{5/2}du\left(\tfrac{u^4}{12}-\tfrac{u^2}{8}+\tfrac{3}{320}\right)\left(\ln\Gamma\left(\tfrac{7}{2}+u\right)
-\ln\Gamma\left(\tfrac{7}{2}-u\right)\right)$&\\
&&\\
8 & $\displaystyle  \frac{1}{90}\zeta_R'(-2)-\frac{1}{72}\zeta_R'(-4)+\frac{1}{360}\zeta_R'(-6)+\ln 8640-\frac{483}{160} $&$\displaystyle 8.7232\times 10^{-2}$ \\
& $\displaystyle -\int_0^{3}du\left(\tfrac{u^5}{60}-\tfrac{u^3}{18}+\tfrac{u}{45}\right)\left(\ln\Gamma\left(4+u\right)
+\ln\Gamma\left(4-u\right)\right)$&\\&&\\
\hline
\hline
\end{tabular}

\end{table}
Using the result of Appendix \ref{a1}, we find that
\begin{equation*}\begin{split}
B^{N}_{\text{ N}\prime}(a;0)=& -\sum_{j=0}^{D-2}y_{D;j}\sum_{i=\left[\frac{j+3}{2}\right]}^N\frac{\left(\tfrac{D-2}{2}\right)^{2i}}{i}\zeta_H\left(2i-j;\tfrac{D}{2}\right)+Y_D\left(\tfrac{2-D}{2}\right),\\
B^{N}_{\text{PC}\prime}(a;0)=& -\sum_{j=0}^{D-2}y_{D;j}\sum_{i=\left[\frac{j+3}{2}\right]}^N\frac{\left(\tfrac{D-2}{2}\right)^{2i}}{i}\zeta_H\left(2i-j;\tfrac{D}{2}\right)+Y_D\left(\tfrac{D-2}{2}\right),\\
B^{N}_{\text{PC}\prime}(a;0)=&-\sum_{j=0}^{D-2}x_{D;j}\sum_{i=\left[\frac{j+3}{2}\right]}^N\frac{\left(\tfrac{D-4}{2}\right)^{2i}}{i}\zeta_H\left(2i-j;\tfrac{D}{2}\right)+Y_D\left(\tfrac{4-D}{2}\right),
\end{split}\end{equation*}where
\begin{align*}
Y_D(c)=&-2z_{D;0}\ln\Gamma\left(\tfrac{D}{2}\right)+\sum_{j=0}^{D-2}z_{D;j}\left\{ (1-(-1)^j)\psi\left(\tfrac{D}{2}\right)\frac{c^{j+1}}{j+1} +c^j\left[ \ln\Gamma\left(\tfrac{D}{2}-c\right)+(-1)^j\ln  \Gamma\left(\tfrac{D}{2}+c\right) \right]\right.\\&\left.
-j\int_0^cdu\,u^{j-1} \left((-1)^j\ln\Gamma\left(\tfrac{D}{2}+u\right)+\ln\Gamma\left(\tfrac{D}{2}-u\right) \right)\right\}.\end{align*}Here $z_{D;j}=y_{D;j}$ for Neumann or perfectly conducting boundary conditions, and $z_{D;j}=x_{D;j}$ for infinitely permeable boundary conditions.

\begin{table}\caption{\label{t7}$\zeta_{\text{PC}}'(a;0)-2\zeta_{\text{PC}}(a;0)\ln a$ }
\begin{tabular}{c|c|c }
\hline
\hline
&&\\
\hspace{0.5cm}$D$\hspace{0.5cm} & \hspace{0.5cm}$\zeta_{\text{PC}}'(a;0)-2\zeta_{\text{PC}}(a;0)\ln a$\hspace{0.5cm} &\hspace{0.5cm} numerical value \hspace{0.5cm}  \\
&&\\
\hline
&&\\
3&   $\displaystyle\frac{3}{8}+2\int_0^{1/2}du\left(\ln\Gamma\left(\tfrac{3}{2}+u\right)
-\ln\Gamma\left(\tfrac{3}{2}-u\right)\right)$ & $\displaystyle 3.8429\times 10^{-1} $ \\
&&\\
4 &$\displaystyle 2\zeta_R'(0)-\zeta_R'(-2)+\ln 2-\frac{3}{2}-2\int_0^{1}du\,u\left(\ln\Gamma\left(2+u\right)
+\ln\Gamma\left(2-u\right)\right)  $&  $\displaystyle-2.9522 $ \\
&& \\
5&$\displaystyle -\frac{13}{12}\zeta_R'(-1)+\frac{7}{12}\zeta_R'(-3)-\frac{4451}{1440}\ln 2 -\frac{3139}{5760}$  &$\displaystyle -1.0179$\\
& $\displaystyle +\int_0^{3/2}du\left(u^2-\tfrac{1}{12}\right)\left(\ln\Gamma\left(\tfrac{5}{2}+u\right)
-\ln\Gamma\left(\tfrac{5}{2}-u\right)\right)$&\\
&& \\
6 & $\displaystyle    \frac{5}{4}\zeta_R'(-2)-\frac{1}{4}\zeta_R'(-4) + \ln 48-\frac{5}{2}
 -\int_0^{2}du\left(\tfrac{u^3}{3}-\tfrac{u}{6}\right)\left(\ln\Gamma\left(3+u\right)+\ln\Gamma\left(3-u\right)\right)$&$\displaystyle   -1.3066 $ \\
&&\\
7 & $\displaystyle   \frac{29}{480}\zeta_R'(-1)-\frac{7}{16}\zeta_R'(-3)+\frac{31}{480}\zeta_R'(-5)-\frac{604127}{12096}\ln 2
  -\frac{7489543}{4838400}$&$\displaystyle -6.5045\times 10^{-1} $   \\
& $\displaystyle +\int_0^{5/2}du\left(\tfrac{u^4}{12}-\tfrac{u^2}{8}+\tfrac{3}{320}\right)\left(\ln\Gamma\left(\tfrac{7}{2}+u\right)
-\ln\Gamma\left(\tfrac{7}{2}-u\right)\right)$&\\
&&\\
8 & $\displaystyle -\frac{5}{36}\zeta_R'(-2)+\frac{11}{72}\zeta_R'(-4)-\frac{1}{72}\zeta_R'(-6)+\ln 4320-\frac{483}{160} $&$\displaystyle -5.9992\times 10^{-1} $ \\
& $\displaystyle -\int_0^{3}du\left(\tfrac{u^5}{60}-\tfrac{u^3}{18}+\tfrac{u}{45}\right)\left(\ln\Gamma\left(4+u\right)
+\ln\Gamma\left(4-u\right)\right)$&\\&&\\
\hline
\hline
\end{tabular}

\end{table}

\begin{table}\caption{\label{t8}$\zeta_{\text{IP}}'(a;0)-2\zeta_{\text{IP}}(a;0)\ln a$ }
\begin{tabular}{c|c|c }
\hline
\hline
&&\\
\hspace{0.5cm}$D$\hspace{0.5cm} & \hspace{0.5cm}$\zeta_{\text{IP}}'(a;0)-2\zeta_{\text{IP}}(a;0)\ln a$\hspace{0.5cm} &\hspace{0.5cm} numerical value \hspace{0.5cm}  \\
&&\\
\hline
&&\\
3&   $\displaystyle \frac{3}{8}+2\int_0^{1/2}du\left(\ln\Gamma\left(\tfrac{3}{2}+u\right)
-\ln\Gamma\left(\tfrac{3}{2}-u\right)\right)$ & $\displaystyle 3.8429\times 10^{-1}$ \\
&&\\
4 &$\displaystyle -2\zeta_R'(0)+\zeta_R'(-2)$ &  $\displaystyle  1.8074$ \\
&& \\
5&$\displaystyle \frac{13}{12}\zeta_R'(-1)-\frac{7}{12}\zeta_R'(-3)+\frac{4451}{1440}\ln 2-\ln3  +\frac{73}{5760}$  &$\displaystyle5.4679\times 10^{-1} $\\
& $\displaystyle +\int_0^{1/2}du\left(3u^2-\tfrac{9}{4}\right)\left(\ln\Gamma\left(\tfrac{5}{2}+u\right)
-\ln\Gamma\left(\tfrac{5}{2}-u\right)\right)$&\\
&& \\
6 & $\displaystyle    -\frac{5}{4}\zeta_R'(-2)+\frac{1}{4}\zeta_R'(-4) - \ln 12+\frac{5}{4}
 -\int_0^{1}du\left(\tfrac{4u^3}{3}-\tfrac{8u}{3}\right)\left(\ln\Gamma\left(3+u\right)+\ln\Gamma\left(3-u\right)\right)$&$\displaystyle 3.6985\times 10^{-1}   $ \\
&&\\
7 & $\displaystyle   -\frac{29}{480}\zeta_R'(-1)+\frac{7}{16}\zeta_R'(-3)-\frac{31}{480}\zeta_R'(-5)+\frac{604127}{120960}\ln 2-\ln 5
 +\frac{3110939}{4838400}$&$\displaystyle  9.8969 \times 10^{-2}$   \\
& $\displaystyle +\int_0^{3/2}du\left(\tfrac{5u^4}{12}-\tfrac{13u^2}{8}+\tfrac{25}{192}\right)\left(\ln\Gamma\left(\tfrac{7}{2}+u\right)
-\ln\Gamma\left(\tfrac{7}{2}-u\right)\right)$&\\
&&\\
8 & $\displaystyle \frac{5}{36}\zeta_R'(-2)-\frac{11}{72}\zeta_R'(-4)+\frac{1}{72}\zeta_R'(-6)-\ln 720+\frac{203}{90} $&$\displaystyle 5.6555\times 10^{-2}$ \\
& $\displaystyle -\int_0^{2}du\left(\tfrac{u^5}{10}-\tfrac{2u^3}{3}+\tfrac{3u}{10}\right)\left(\ln\Gamma\left(4+u\right)
+\ln\Gamma\left(4-u\right)\right)$&\\&&\\
\hline
\hline
\end{tabular}
\end{table}

Taking the derivatives of \eqref{eq4_10_1}   at $s=0$, we find that
\begin{equation}\label{eq4_16_2}\begin{split}
\zeta_{\text{D}}'(a;0)-2\zeta_{\text{D}}(a;0)\ln a=&- \sum_{j=0}^{D-2}y_{D;j}\zeta_H'\left(-j;\tfrac{D-2}{2}\right)
-\sum_{i=1}^{\left[\tfrac{D-1}{2}\right]}y_{D;2i-1}\sum_{k=0}^{2i}d_{2i,k}\left(\psi(i+k)-\psi(1)-2\psi\left(\tfrac{D-2}{2}\right)\right),\\
\zeta_{\text{N}}'(a;0)-2\zeta_{\text{N}}(a;0)\ln a=& \sum_{j=0}^{D-2}y_{D;j}\zeta_H'\left(-j;\tfrac{D}{2}\right)+ \sum_{j=0}^{D-2}\sum_{i=1}^{\left[\tfrac{j}{2}\right]} y_{D;j}\frac{\left(\tfrac{D-2}{2}\right)^{2i}}{i}\zeta_H\left(2i-j;\tfrac{D}{2}\right)+Y_D\left(\tfrac{2-D}{2}\right)+\ln\tfrac{D}{2}\\&-
\sum_{i=1}^{\left[\tfrac{D-1}{2}\right]}y_{D;2i-1}\sum_{k=0}^{2i} m_{2i,k}\left(\tfrac{2-D}{2}\right) \left(\psi(i+k)-\psi(1)-2\psi\left(\tfrac{D}{2}\right)\right),\\
\zeta_{\text{PC}}'(a;0)-2\zeta_{\text{PC}}(a;0)\ln a=&- \sum_{j=0}^{D-2}(x_{D;j}-y_{D;j})\zeta_H'\left(-j;\tfrac{D}{2}\right)
+ \sum_{j=0}^{D-2}\sum_{i=1}^{\left[\tfrac{j}{2}\right]} y_{D;j}\frac{\left(\tfrac{D-2}{2}\right)^{2i}}{i}\zeta_H\left(2i-j;\tfrac{D}{2}\right)+Y_D\left(\tfrac{D-2}{2}\right)\\&-\sum_{i=1}^{\left[\tfrac{D-1}{2}\right]}\sum_{k=0}^{2i}
\left(x_{D;2i-1}d_{2i,k}+y_{D;2i-1}m_{2i,k}\left(\tfrac{D-2}{2}\right)\right)\left(\psi(i+k)-\psi(1)-2\psi\left(\tfrac{D}{2}\right)\right),\end{split}\end{equation}
\begin{equation}\label{eq4_16_2_2}\begin{split}
\zeta_{\text{IP}}'(a;0)-2\zeta_{\text{IP}}(a;0)\ln a=& \sum_{j=0}^{D-2}(x_{D;j}-y_{D;j})\zeta_H'\left(-j;\tfrac{D}{2}\right)
+ \sum_{j=0}^{D-2}\sum_{i=1}^{\left[\tfrac{j}{2}\right]} x_{D;j}\frac{\left(\tfrac{D-4}{2}\right)^{2i}}{i}\zeta_H\left(2i-j;\tfrac{D}{2}\right)+Y_D\left(\tfrac{4-D}{2}\right)\\&-\sum_{i=1}^{\left[\tfrac{D-1}{2}\right]}\sum_{k=0}^{2i}
\left(x_{D;2i-1}m_{2i,k}\left(\tfrac{4-D}{2}\right)+y_{D;2i-1}d_{2i,k}\right)\left(\psi(i+k)-\psi(1)-2\psi\left(\tfrac{D}{2}\right)\right).
\end{split}\end{equation}

The values of  $\zeta'(a;0)-2\zeta(a;0)\ln a$ are computed explicitly in Tables \ref{t5}, \ref{t6}, \ref{t7} and \ref{t8} for $3\leq D\leq 8$.

\section{The renormalized Casimir energy}

In this section, we derive the expressions for the renormalized Casimir free energies. First we consider the functions $\Xi(s,\alpha;\chi;c)$ and $X(s)$ defined in \eqref{eq4_22_1} and \eqref{eq4_22_2}. We have
\begin{align*}
\Xi(s,\alpha;\chi;c)=&\sum_{n=0}^{\infty}\sum_{p=-\infty}^{\infty}\frac{(n+\chi)^{\alpha}}{\left((n+\chi)^2+(cp)^2\right)^{s}}\\
=& \sum_{n=0}^{\infty}\sum_{p=-\infty}^{\infty}\frac{(n+\chi)^{\alpha}}{\Gamma(s)}\int_0^{\infty}dt\, t^{s-1}\exp\left\{-t\left([n+\chi]^2+[cp]^2\right)\right\}\\
=&\sum_{n=0}^{\infty}\sum_{p=-\infty}^{\infty}\frac{(n+\chi)^{\alpha}}{\Gamma(s)}\frac{\sqrt{\pi}}{c}\int_0^{\infty}dt\, t^{s-1-1/2}\exp\left\{-t[n+\chi]^2-\frac{\pi^2p^2}{tc^2}\right\}\\
=&\frac{\sqrt{\pi}}{c}\frac{\Gamma\left(s-\tfrac{1}{2}\right)}{\Gamma(s)}\zeta_H(2s-1-\alpha;\chi)+\frac{4\sqrt{\pi}}{c\Gamma(s)}\sum_{n=0}^{\infty}\sum_{p=1}^{\infty}
(n+\chi)^{\alpha-s+1/2}\left(\frac{\pi p}{c}\right)^{s-1/2}K_{s-1/2}\left(\frac{2\pi p}{c}(n+\chi)\right)\\
=&\Xi_{\text{singular}}(s,\alpha;\chi;c)+\Xi_{\text{regular}}(s,\alpha;\chi;c),
\end{align*}
where
\begin{align*}
\Xi_{\text{singular}}(s,\alpha;\chi;c)=&\frac{\sqrt{\pi}}{c}\frac{\Gamma\left(s-\tfrac{1}{2}\right)}{\Gamma(s)}\zeta_H(2s-1-\alpha;\chi)
\end{align*}is the singular part of $\Xi(s,\alpha;\chi;c)$ with poles at $s=1+\alpha/2, 1/2, -1/2, -3/2, \ldots$, and
\begin{align*}
\Xi_{\text{regular}}(s,\alpha;\chi;c)=&\frac{4\sqrt{\pi}}{c\Gamma(s)}\sum_{n=0}^{\infty}\sum_{p=1}^{\infty}
(n+\chi)^{\alpha-s+1/2}\left(\frac{\pi p}{c}\right)^{s-1/2}K_{s-1/2}\left(\frac{2\pi p}{c}(n+\chi)\right)
\end{align*}is a regular function.

\begin{table}\caption{\label{t9}$E^{\text{asym}}$ for Dirichlet and Neumann boundary conditions }
\begin{tabular}{c|c|c   }
\hline
\hline
&&\\
\hspace{0.5cm}$D$\hspace{0.5cm} & \hspace{0.5cm}Dirichlet\hspace{0.5cm} &\hspace{0.5cm}Neumann\hspace{0.5cm}  \\
&&\\
\hline
&&\\
3&   $\displaystyle \frac{1}{24}T\ln(aT)+0.1476 T$ & $\displaystyle \frac{17}{24}T\ln(aT)+0.3041 T $ \\
&&\\
4 &$\displaystyle -0.0152 T-0.0008 \ln (aT) -0.0015$ &  $\displaystyle  T\ln(aT)+0.2410T+0.1265\ln(aT)+0.2471$ \\
&& \\
5&$\displaystyle -\frac{17}{5760}T\ln(aT)-0.0080 T$  &$\displaystyle \frac{3887}{5760}T\ln(aT)+0.0998 T $\\
&&\\
6 & $\displaystyle  0.0016 T+0.0001\ln (aT)+0.0002   $&$\displaystyle T\ln(aT)+0.0824T+0.1453\ln(aT)+0.2839  $ \\
&&\\
7 & $\displaystyle \frac{367}{967680}T\ln(aT)+0.0010 T  $&$\displaystyle  \frac{676463}{967680}T\ln (aT)+0.0114 T$   \\
&&\\
8 & $\displaystyle  -0.00023 T-0.00001 \ln (aT)-0.00003 $&$\displaystyle T\ln (aT)-0.0436T+ 0.1568\ln(aT)+0.3063$ \\
 &&\\
\hline
\hline
\end{tabular}
\end{table}

\begin{table}\caption{\label{t10}$E^{\text{asym}}$ for perfectly conducting and infinitely permeable boundary conditions }
\begin{tabular}{c|c|c  }
\hline
\hline
&&\\
\hspace{0.5cm}$D$\hspace{0.5cm}   & \hspace{0.5cm}perfectly conducting\hspace{0.5cm} &\hspace{0.5cm}infinitely permeable\hspace{0.5cm}\\
&&\\
\hline
&&\\
3&   $\displaystyle -\frac{1}{ 4}T\ln(aT)-0.1921 T$ & $\displaystyle -\frac{1}{ 4}T\ln(aT)-0.1921 T $ \\
&&\\
4 &$\displaystyle T\ln (aT)+1.4761 T+ 0.0352 \ln (aT) + 0.0688$ &  $\displaystyle  -T\ln (aT)-0.9037 T -0.0585 \ln (aT)  -0.1144$ \\
&& \\
5&$\displaystyle \frac{899}{1440}T\ln(aT)+ 0.5090 T$  &$\displaystyle-\frac{989}{1440}T\ln(aT)-0.2734 T $\\
&&\\
6 & $\displaystyle  T\ln(aT)+0.6533 T+0.0888\ln (aT)+0.1734   $&$\displaystyle -T\ln(aT)-0.1849 T-0.1469\ln (aT)-0.2870  $ \\
&&\\
7 & $\displaystyle \frac{340577}{483840}T\ln(aT)+0.3252 T  $&$\displaystyle  -\frac{301727}{483840}T\ln(aT) -0.0495T$   \\
&&\\
8 & $\displaystyle  T\ln(aT)+0.3000 T+0.1117 \ln (aT)+0.2183 $&$\displaystyle -T\ln(aT)-0.0283 T-0.1745 \ln (aT) -0.3409$ \\
 &&\\
\hline
\hline
\end{tabular}
\end{table}

Similarly,
\begin{align*}
X(s)=&\sum_{p=-\infty}^{\infty}\frac{1}{\left(\left[\tfrac{D}{2}\right]^2+[2\pi p aT]^2\right)^{s}}\\
=&\sum_{p=-\infty}^{\infty}\frac{1}{\Gamma(s)}\int_0^{\infty} dt\, t^{s-1}\exp\left\{-t\left(\left[\tfrac{D}{2}\right]^2+[2\pi p aT]^2\right)\right\}\\
=&\sum_{p=-\infty}^{\infty}\frac{1}{\Gamma(s)}\frac{\sqrt{\pi}}{2\pi aT}\int_0^{\infty} dt\, t^{s-1-1/2}\exp\left\{-t\left[\tfrac{D}{2}\right]^2-\frac{p^2}{4ta^2T^2}\right\}\\
=&\frac{\Gamma\left(s-1/2\right)}{\Gamma(s)} \frac{1}{2\sqrt{\pi} aT}\left(\frac{2}{D}\right)^{2s-1}+\frac{2}{\Gamma(s)} \frac{1}{ \sqrt{\pi} aT}\sum_{p=1}^{\infty}
\left(\frac{p}{DaT}\right)^{s-1/2}K_{s-1/2}\left(\frac{Dp}{2aT}\right)\\
=&X_{\text{singular}}(s)+X_{\text{regular}}(s),
\end{align*}
where
$$X_{\text{singular}}(s)=\frac{\Gamma\left(s-1/2\right)}{\Gamma(s)} \frac{1}{2\sqrt{\pi} aT}\left(\frac{2}{D}\right)^{2s-1}$$ has poles at $s=1/2, -1/2, -3/2, \ldots$ and $$X_{\text{regular}}(s)= \frac{2}{\Gamma(s)}\frac{1}{ \sqrt{\pi} aT}\sum_{p=1}^{\infty}
\left(\frac{p}{DaT}\right)^{s-1/2}K_{s-1/2}\left(\frac{Dp}{2aT}\right)$$ is a regular function.

Substituting these into \eqref{eq4_22_3}, we  can compute $\zeta_T'(a;0)$. Using \eqref{eq4_23_1} and \eqref{eq4_23_2}, one can then write down the expressions for the renormalized Casimir free energy. We find that  for Dirichlet boundary conditions:
\begin{align*}
E_{\text{Cas}}^{\text{ren}}=& -\frac{1}{4a }\sum_{j=0}^{D-2}y_{D;j}\zeta_H\left(-j-1;\tfrac{D-2}{2}\right) +\frac{1}{2\sqrt{\pi} a}
\sum_{j=0}^{D-2}y_{D;j}\sum_{n=0}^{\infty}\sum_{p=1}^{\infty}\left(n+\tfrac{D-2}{2}\right)^{j+1/2}\sqrt{\frac{2aT}{ p}} K_{ 1/2}\left(\frac{ p}{aT}\left(n+\tfrac{D-2}{2}\right)\right)\\
&+\frac{1}{2\sqrt{\pi}a}\sum_{i=1}^{\left[\tfrac{D}{2}\right]}\sum_{\substack{j=0\\j\neq 2i-2}}^{D-2}\sum_{k=0}^{2i}y_{D;j}d_{2i,k}\frac{\Gamma\left(i+k-\tfrac{1}{2}\right)}
{\Gamma(i+k)}\zeta_H\left(2i-j-1;\tfrac{D-2}{2}\right)\\&+\frac{1}{4\sqrt{\pi}a}\sum_{i=1}^{\left[\tfrac{D}{2}\right]}
\sum_{k=0}^{2i}y_{D;2i-2}d_{2i,k}\frac{\Gamma\left(i+k-\frac{1}{2}\right)}{\Gamma(i+k)}\left(\psi\left(i+k-\tfrac{1}{2}\right)-\psi(1)-2\psi\left(\tfrac{D-2}{2}\right)\right)\\
&+\frac{2}{\sqrt{\pi}a  }\sum_{j=0}^{D-2}y_{D;j}\sum_{i=1}^{\left[\tfrac{D}{2}\right]}\sum_{k=0}^{2i}\frac{d_{2i,k}}{\Gamma(i+k)}\sum_{n=0}^{\infty}\sum_{p=1}^{\infty}
\left(n+\tfrac{D-2}{2}\right)^{k+j-i+1/2}\left(\frac{ p}{2aT}\right)^{i+k-1/2}K_{i+k-1/2}\left(\frac{  p}{aT}\left(n+\tfrac{D-2}{2}\right)\right)\\
&-\frac{T}{2}\sum_{l=0}^{\infty}\sum_{p=-\infty}^{\infty}b_D(l)B_{\text{D},\left[\frac{D}{2}\right]}^{ l+\frac{D-2}{2}\prime}(a;0;2\pi p T)-\frac{\hat{c}_{D+1}}{2\sqrt{\pi}}\ln(a\mu)+\frac{1}{\sqrt{\pi}}\sum_{n=0}^{D-1}2^{D-n} \Gamma\left(\frac{D-n+1}{2}\right)\zeta_R(D-n+1) \hat{c}_{n} T^{D-n+1}.
\end{align*}
\begin{figure}
\epsfxsize=0.43\linewidth \epsffile{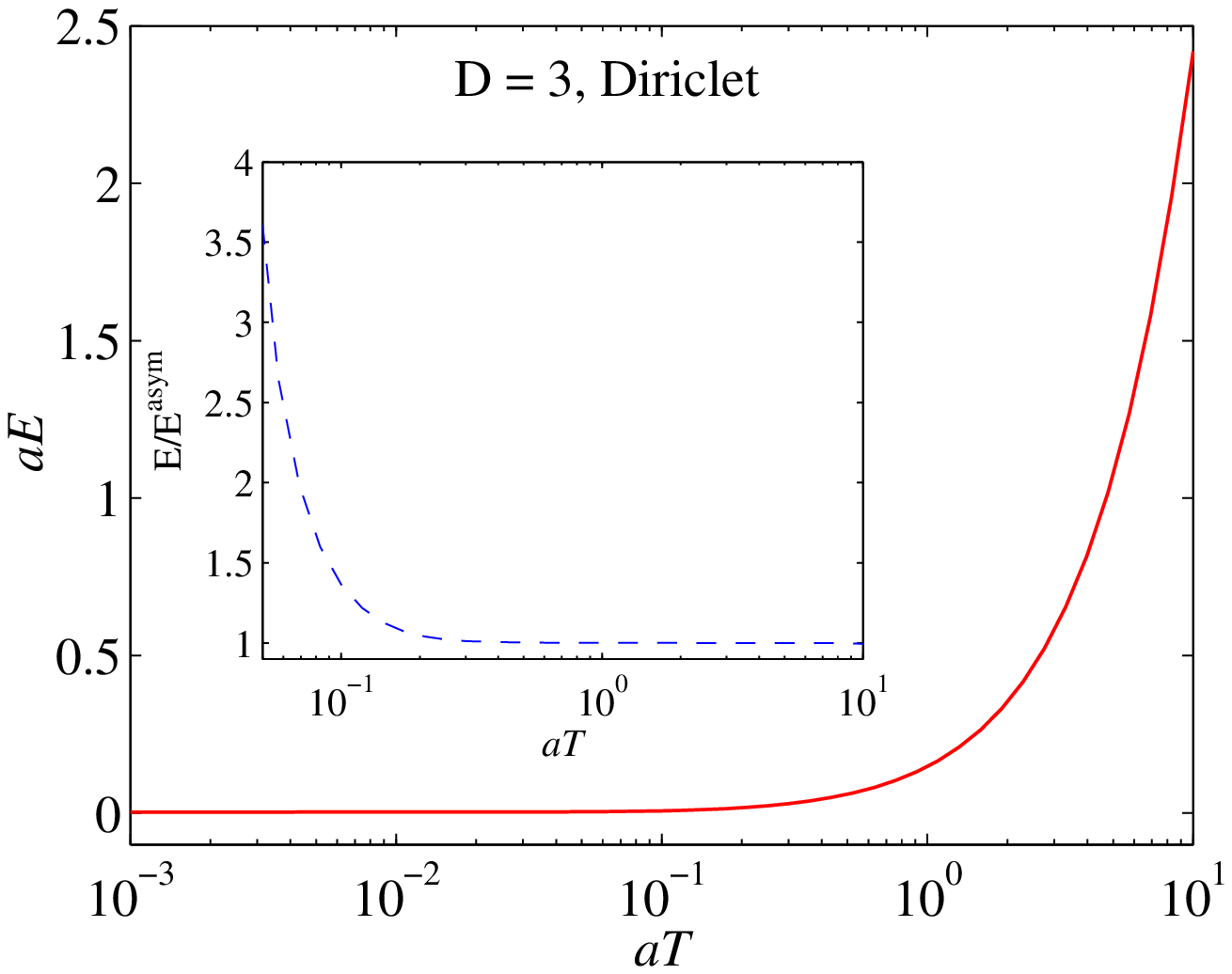}  \epsfxsize=0.43\linewidth \epsffile{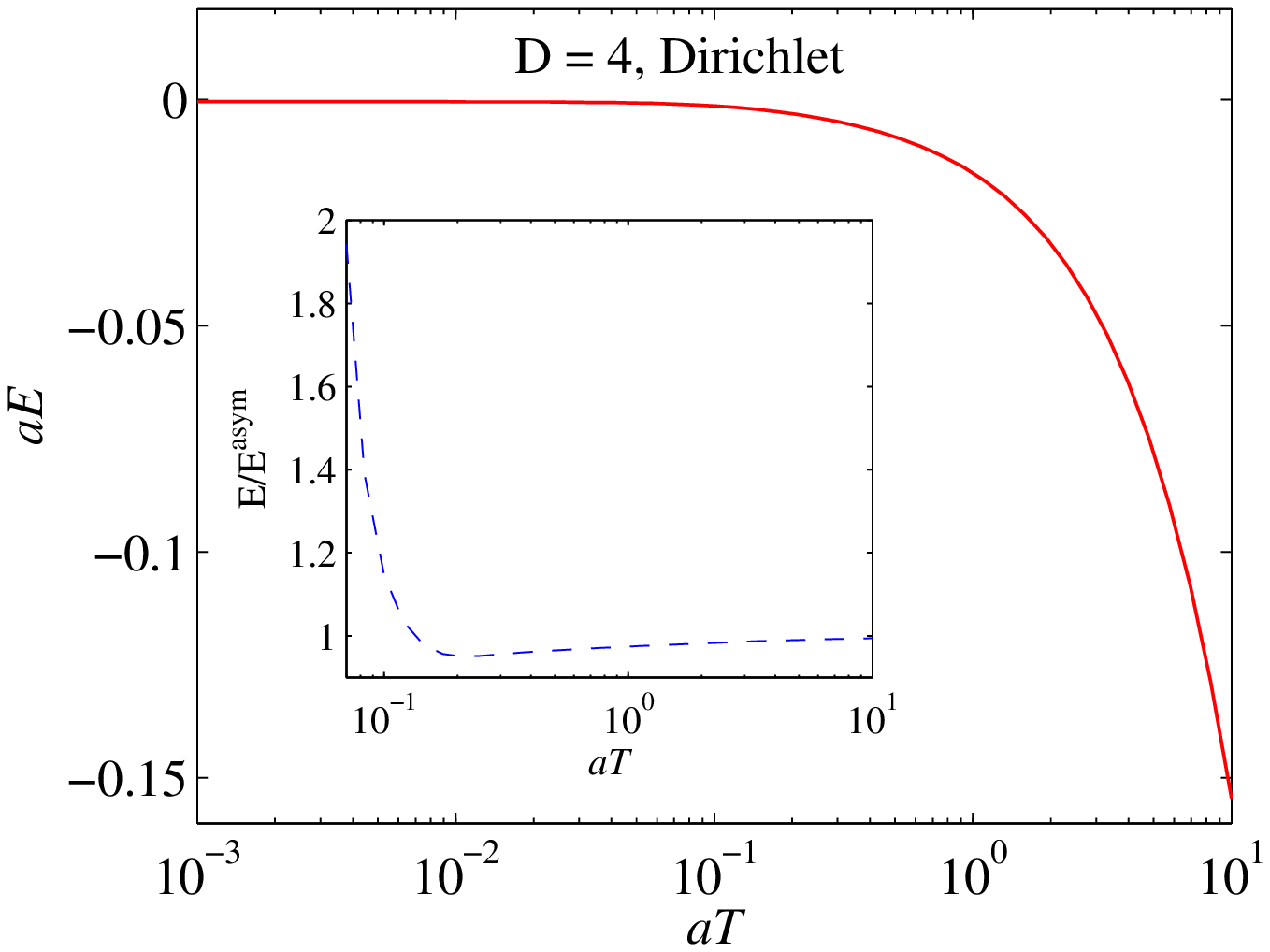}\caption{\label{f1} $aE$  as a function of $aT$ for Dirichlet boundary conditions when $D=3$ and $D=4$. The inset shows the ratio of $E$ to $E^{\text{asym}}$. }\end{figure}

\begin{figure}
\epsfxsize=0.43\linewidth \epsffile{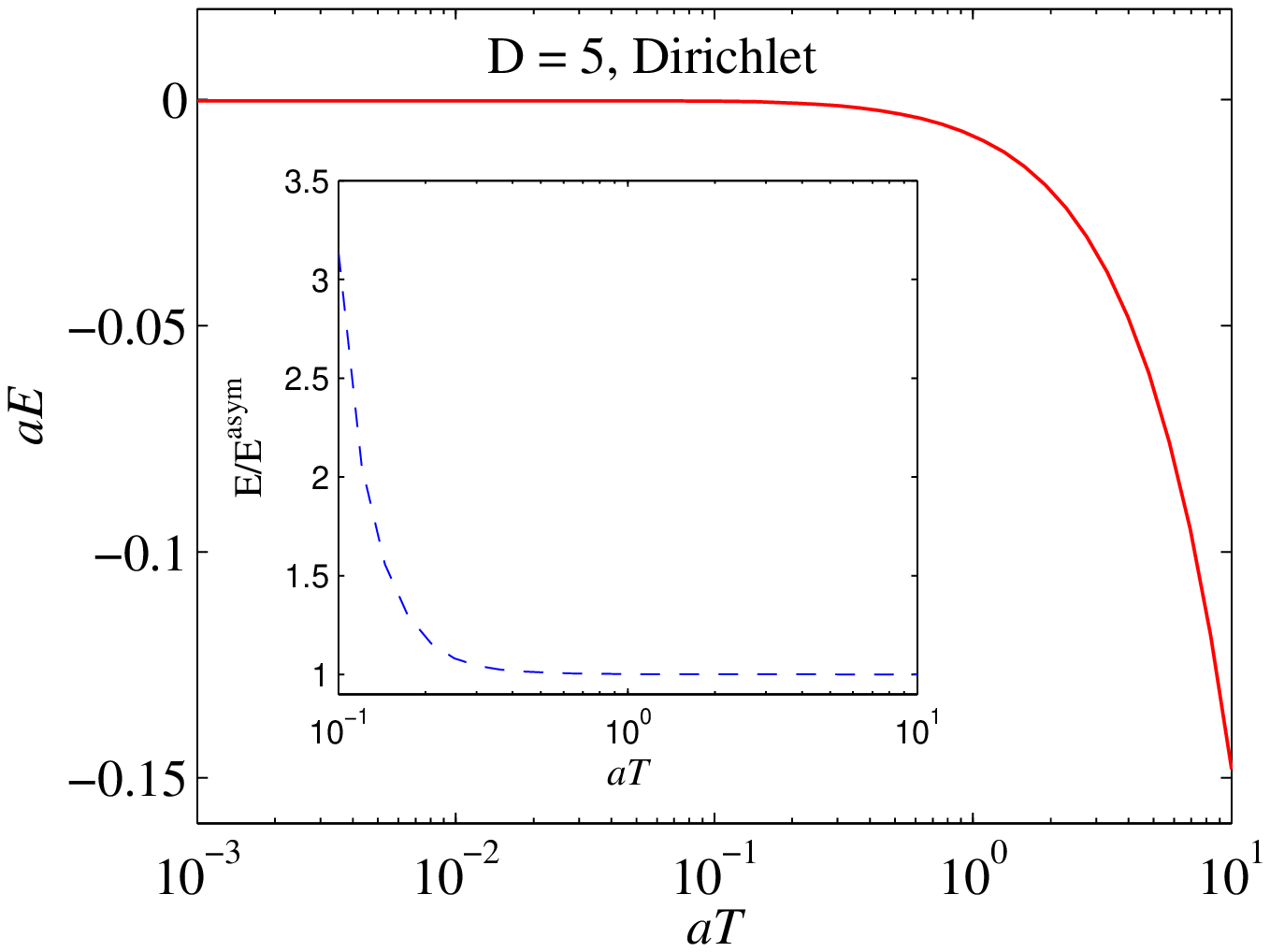}  \epsfxsize=0.43\linewidth \epsffile{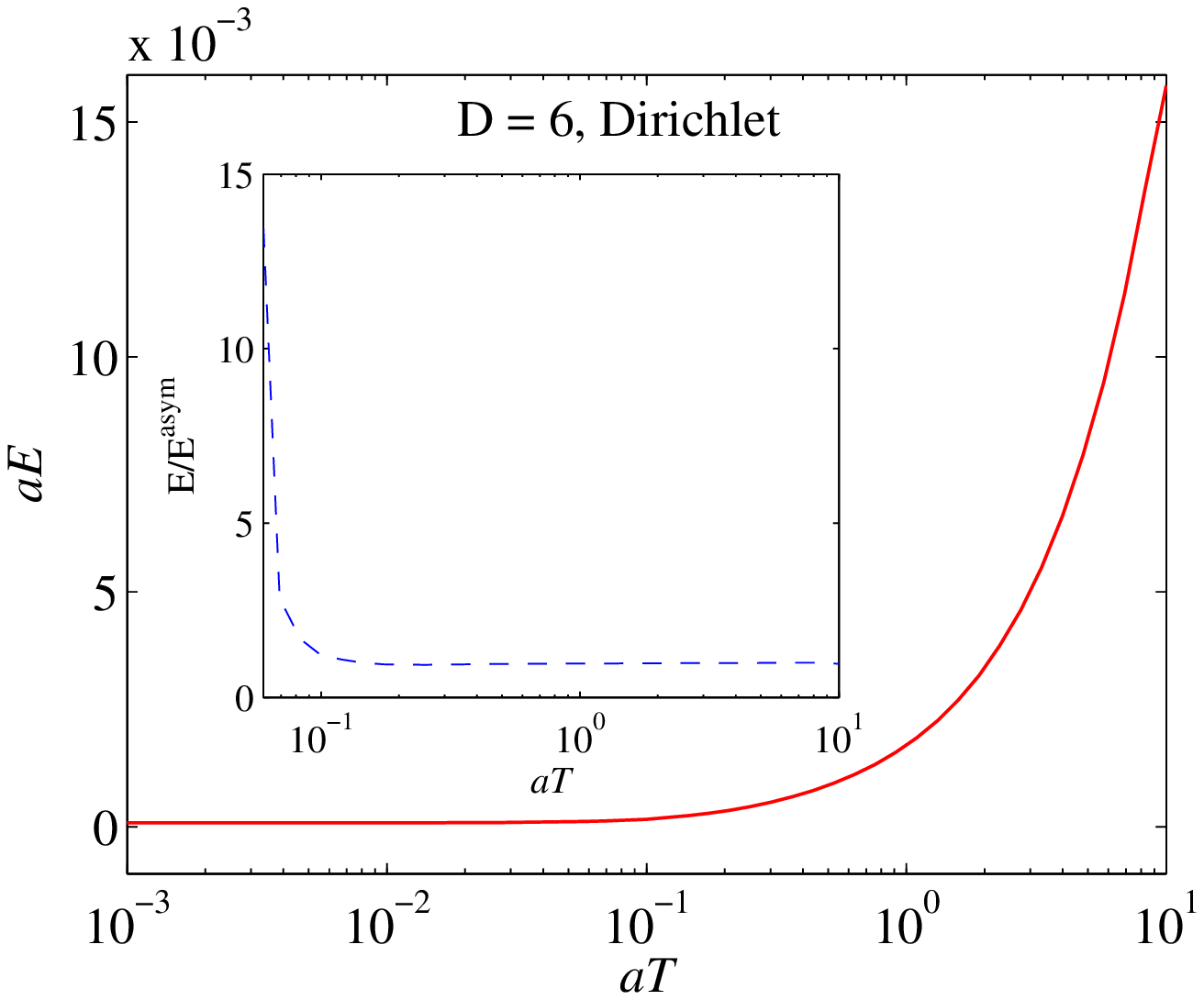}\caption{\label{f2} $aE$  as a function of $aT$ for Dirichlet boundary conditions when $D=5$ and $D=6$. The inset shows the ratio of $E$ to $E^{\text{asym}}$. }\end{figure}
For Neumann boundary conditions:
\begin{align*}
E_{\text{Cas}}^{\text{ren}}=&  \frac{1}{4a}\sum_{j=0}^{D-2}y_{D;j}\zeta_H\left(-j-1;\tfrac{D}{2}\right) -\frac{1}{2\sqrt{\pi} a}
\sum_{j=0}^{D-2}y_{D;j}\sum_{n=0}^{\infty}\sum_{p=1}^{\infty}\left(n+\tfrac{D}{2}\right)^{j+1/2}\sqrt{\frac{2aT}{ p}} K_{ 1/2}\left(\frac{ p}{aT}\left(n+\tfrac{D}{2}\right)\right)\\
&+\frac{1}{2\sqrt{\pi}a}\sum_{i=1}^{\left[\tfrac{D}{2}\right]}\sum_{\substack{j=0\\j\neq 2i-2}}^{D-2}\sum_{k=0}^{2i}y_{D;j}m_{2i,k}\left(\tfrac{2-D}{2}\right)\frac{\Gamma\left(i+k-\tfrac{1}{2}\right)}
{\Gamma(i+k)}\zeta_H\left(2i-j-1;\tfrac{D}{2}\right)\\
&+\frac{1}{4\sqrt{\pi}a}\sum_{i=1}^{\left[\tfrac{D}{2}\right]}
\sum_{k=0}^{2i}y_{D;2i-2}m_{2i,k}\left(\tfrac{2-D}{2}\right)\frac{\Gamma\left(i+k-\frac{1}{2}\right)}{\Gamma(i+k)}\left(\psi\left(i+k-\tfrac{1}{2}\right)-\psi(1)-2\psi\left(\tfrac{D}{2}\right)\right)\\
&+\frac{2}{\sqrt{\pi}a}\sum_{j=0}^{D-2}y_{D;j}\sum_{i=1}^{\left[\tfrac{D}{2}\right]}\sum_{k=0}^{2i}\frac{m_{2i,k}\left(\tfrac{2-D}{2}\right)}{\Gamma(i+k)}\sum_{n=0}^{\infty}\sum_{p=1}^{\infty}
\left(n+\tfrac{D}{2}\right)^{k+j-i+1/2}\left(\frac{ p}{2aT}\right)^{i+k-1/2}K_{i+k-1/2}\left(\frac{  p}{aT}\left(n+\tfrac{D}{2}\right)\right)\\
&-\frac{T}{2}\sum_{l=1}^{\infty}\sum_{p=-\infty}^{\infty}b_D(l)B_{\left(\text{R},\tfrac{2-D}{2}\right),\left[\frac{D}{2}\right]}^{ l+\frac{D-2}{2}\prime}(a;0;2\pi p T) -\frac{D}{8  a} +\frac{1}{ 2\sqrt{\pi} a }\sum_{p=1}^{\infty}
\sqrt{\frac{DaT}{p}}K_{1/2}\left(\frac{Dp}{2aT}\right)-\frac{T}{2}\sum_{p=-\infty}^{\infty}
B_{D,1}^{\tfrac{D}{2}\prime}(a;0;2\pi pT)\\&+\frac{1}{\sqrt{\pi}a}\sum_{k=0}^2d_{2,k}\left(\tfrac{D}{2}\right)^{2k}\left(\frac{\Gamma\left(k+1/2\right)}{\Gamma(k+1)} \frac{1}{2 }\left(\frac{2}{D}\right)^{2k+1}+\frac{2}{\Gamma(k+1)} \sum_{p=1}^{\infty}
\left(\frac{p}{DaT}\right)^{k+1/2}K_{k+1/2}\left(\frac{Dp}{2aT}\right)\right)\\&-\frac{\hat{c}_{D+1}}{2\sqrt{\pi}}\ln(a\mu)+\frac{1}{\sqrt{\pi}}\sum_{n=0}^{D-1}2^{D-n} \Gamma\left(\frac{D-n+1}{2}\right)\zeta_R(D-n+1) \hat{c}_{n} T^{D-n+1}.
\end{align*}
\begin{figure}
\epsfxsize=0.43\linewidth \epsffile{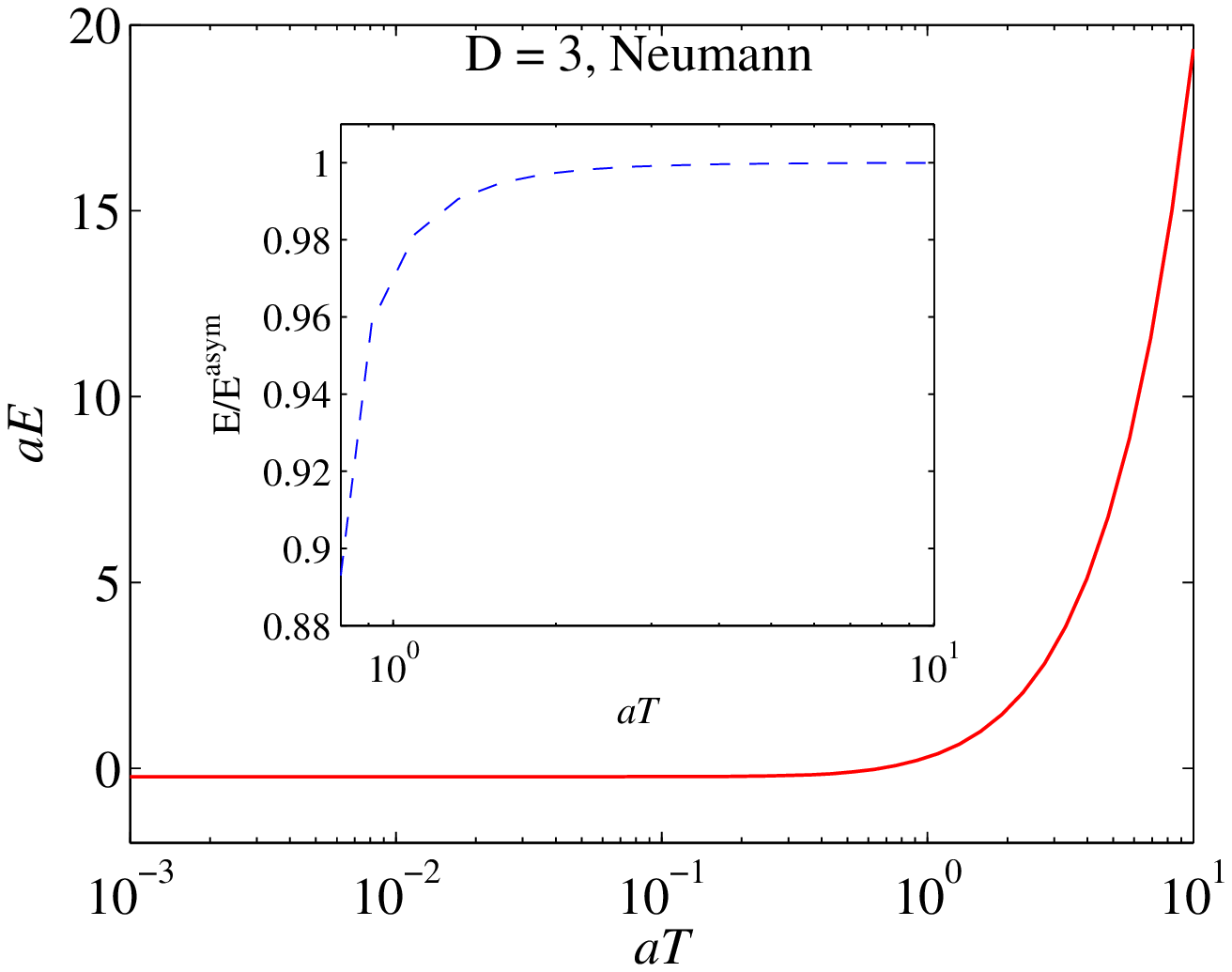}  \epsfxsize=0.43\linewidth \epsffile{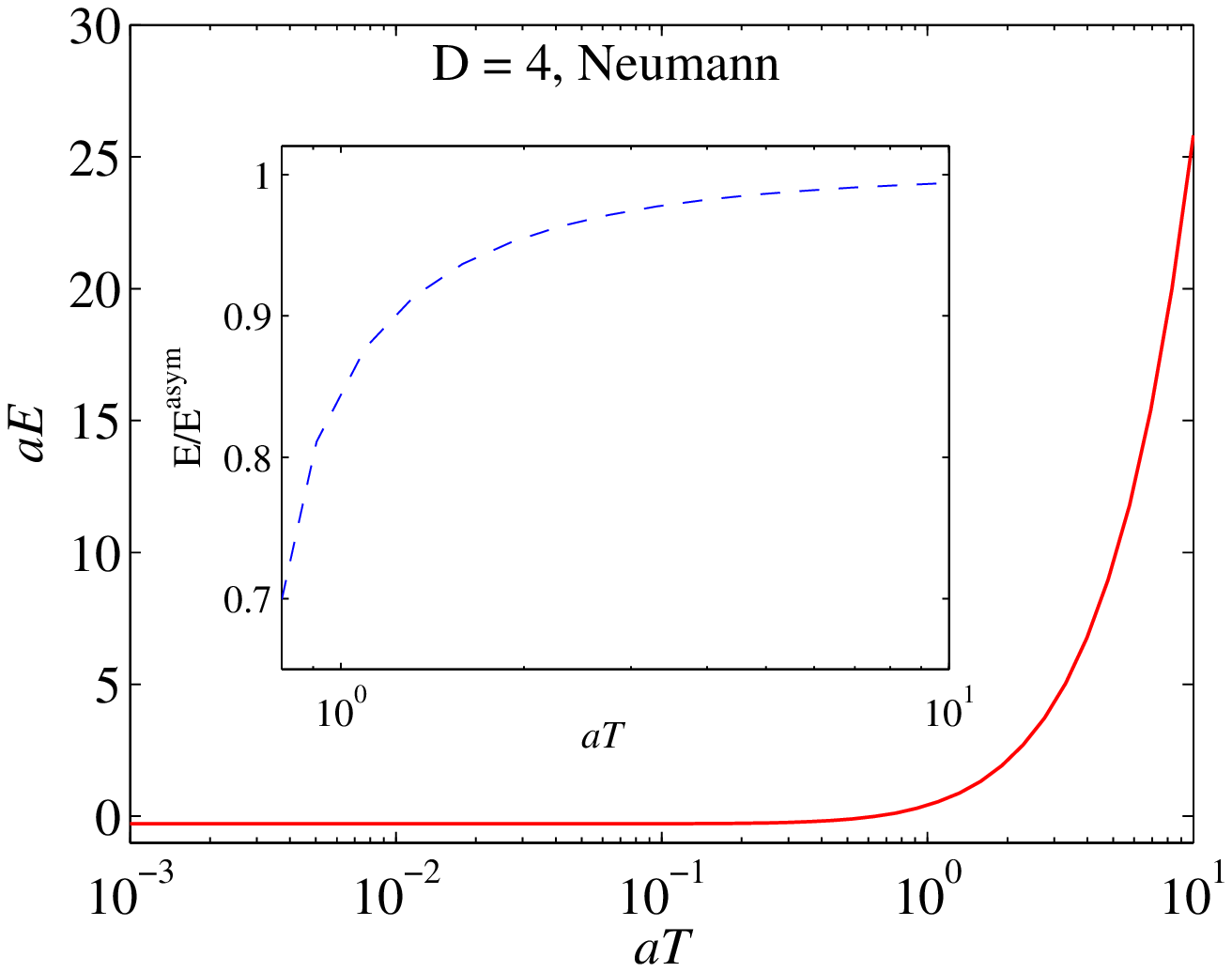}\caption{\label{f3} $aE$  as a function of $aT$ for Neumann boundary conditions when $D=3$ and $D=4$. The inset shows the ratio of $E$ to $E^{\text{asym}}$. }\end{figure}

\begin{figure}
\epsfxsize=0.43\linewidth \epsffile{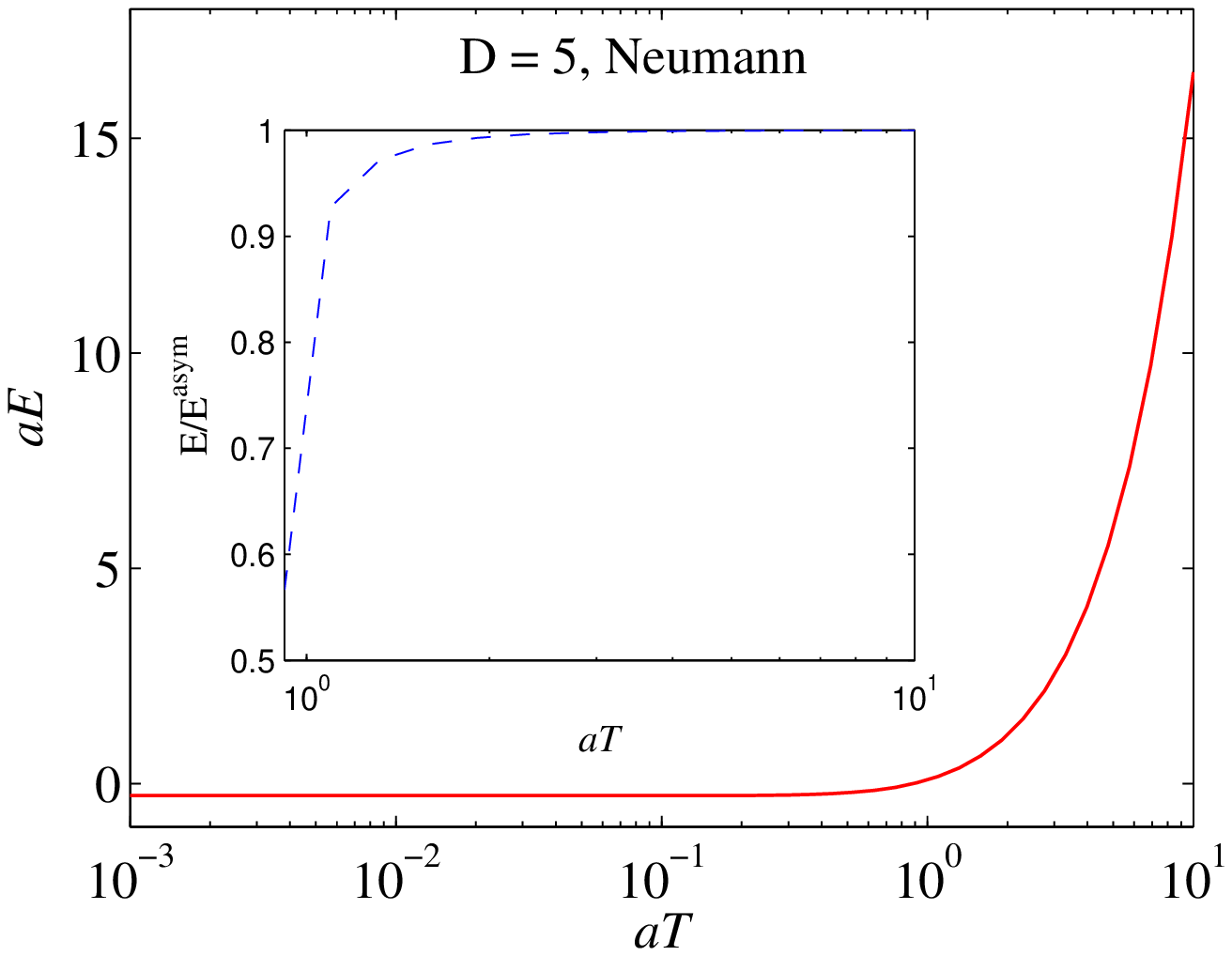}  \epsfxsize=0.43\linewidth \epsffile{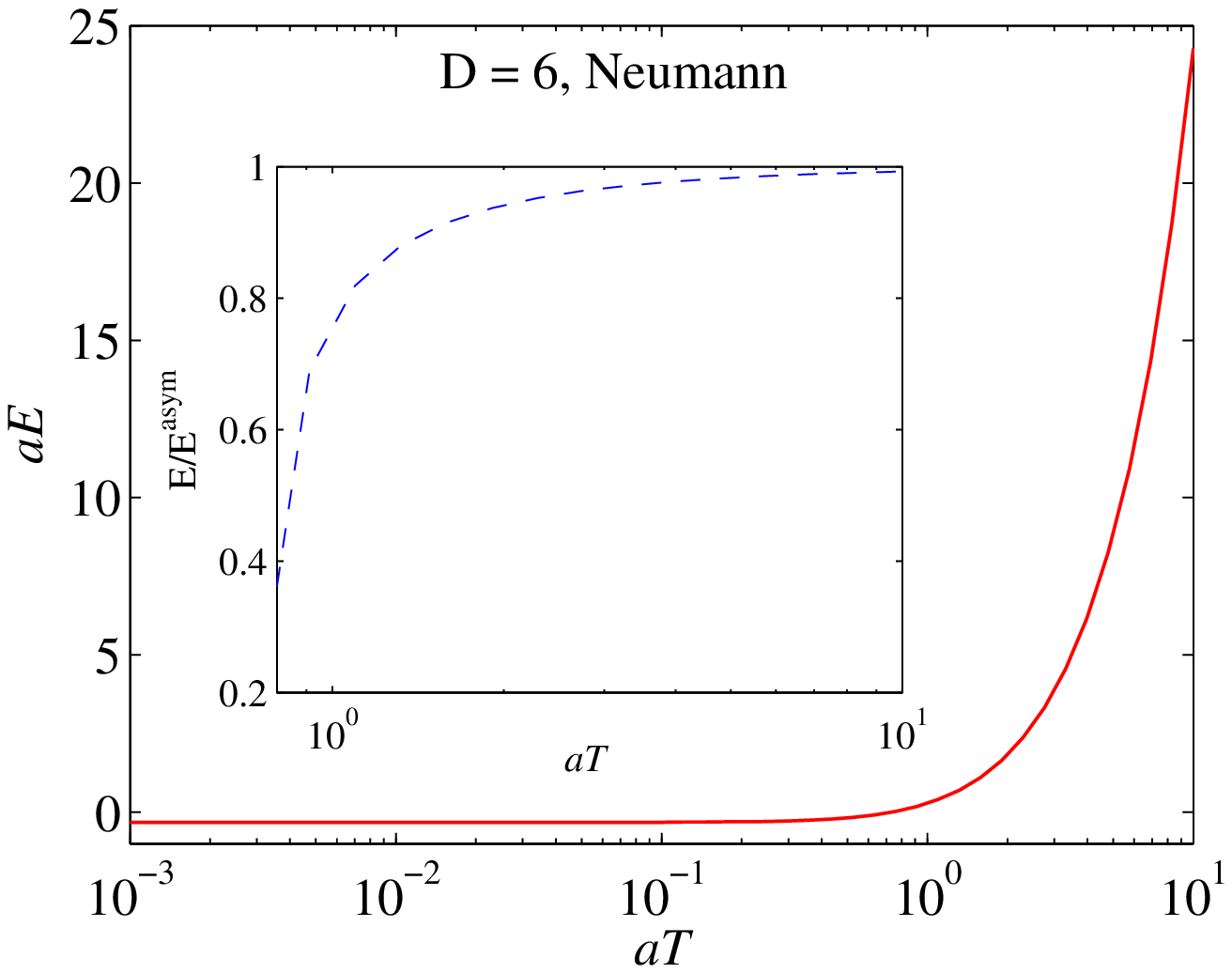}\caption{\label{f4} $aE$  as a function of $aT$ for Neumann boundary conditions when $D=5$ and $D=6$. The inset shows the ratio of $E$ to $E^{\text{asym}}$. }\end{figure}For perfectly conducting boundary conditions:
\begin{align*}
E_{\text{Cas}}^{\text{ren}}=&  - \frac{1}{4a}\sum_{j=0}^{D-2}\left(x_{D;j}-y_{D;j}\right)\zeta_H\left(-j-1;\tfrac{D}{2}\right) +\frac{1}{2\sqrt{\pi} a}
\sum_{j=0}^{D-2}\left(x_{D;j}-y_{D;j}\right)\sum_{n=0}^{\infty}\sum_{p=1}^{\infty}\left(n+\tfrac{D}{2}\right)^{j+1/2}\sqrt{\frac{2aT}{ p}} K_{ 1/2}\left(\frac{ p}{aT}\left(n+\tfrac{D}{2}\right)\right)\\
&+\frac{1}{2\sqrt{\pi}a}\sum_{i=1}^{\left[\tfrac{D}{2}\right]}\sum_{\substack{j=0\\j\neq 2i-2}}^{D-2}\sum_{k=0}^{2i}\left(x_{D;j}d_{2i,k}+y_{D;j}m_{2i,k}\left(\tfrac{D-2}{2}\right)\right)\frac{\Gamma\left(i+k-\tfrac{1}{2}\right)}
{\Gamma(i+k)}\zeta_H\left(2i-j-1;\tfrac{D}{2}\right)\\&+\frac{1}{4\sqrt{\pi}a}\sum_{i=1}^{\left[\tfrac{D}{2}\right]}
\sum_{k=0}^{2i}\left(x_{D;2i-2}d_{2i,k}+y_{D;2i-2}m_{2i,k}\left(\tfrac{D-2}{2}\right)\right)
\frac{\Gamma\left(i+k-\frac{1}{2}\right)}{\Gamma(i+k)}\left(\psi\left(i+k-\tfrac{1}{2}\right)-\psi(1)-2\psi\left(\tfrac{D}{2}\right)\right)\\
&+\frac{2}{\sqrt{\pi}a }\sum_{j=0}^{D-2} \sum_{i=1}^{\left[\tfrac{D}{2}\right]}\sum_{k=0}^{2i}\frac{\left(x_{D;j}d_{2i,k}+y_{D;j}m_{2i,k}\left(\tfrac{D-2}{2}\right)\right)}
{\Gamma(i+k)}\sum_{n=0}^{\infty}\sum_{p=1}^{\infty}
\left(n+\tfrac{D}{2}\right)^{k+j-i+1/2}\left(\frac{ p}{2aT}\right)^{i+k-1/2}K_{i+k-1/2}\left(\frac{  p}{aT}\left(n+\tfrac{D}{2}\right)\right)\\
&-\frac{T}{2}\sum_{l=1}^{\infty}\sum_{p=-\infty}^{\infty}\left(
h_D(l)B_{ \text{D},\left[\frac{D}{2}\right]}^{ l+\frac{D-2}{2}\prime}(a;0;2\pi p T)+b_D(l)B_{\left(\text{R},\tfrac{D-2}{2}\right),\left[\frac{D}{2}\right]}^{ l+\frac{D-2}{2}\prime}(a;0;2\pi p T)\right)\\&-\frac{\hat{c}_{D+1}}{2\sqrt{\pi}}\ln(a\mu)+\frac{1}{\sqrt{\pi}}\sum_{n=0}^{D-1}2^{D-n} \Gamma\left(\frac{D-n+1}{2}\right)\zeta_R(D-n+1) \hat{c}_{n} T^{D-n+1}.
\end{align*}
\begin{figure}
\epsfxsize=0.43\linewidth \epsffile{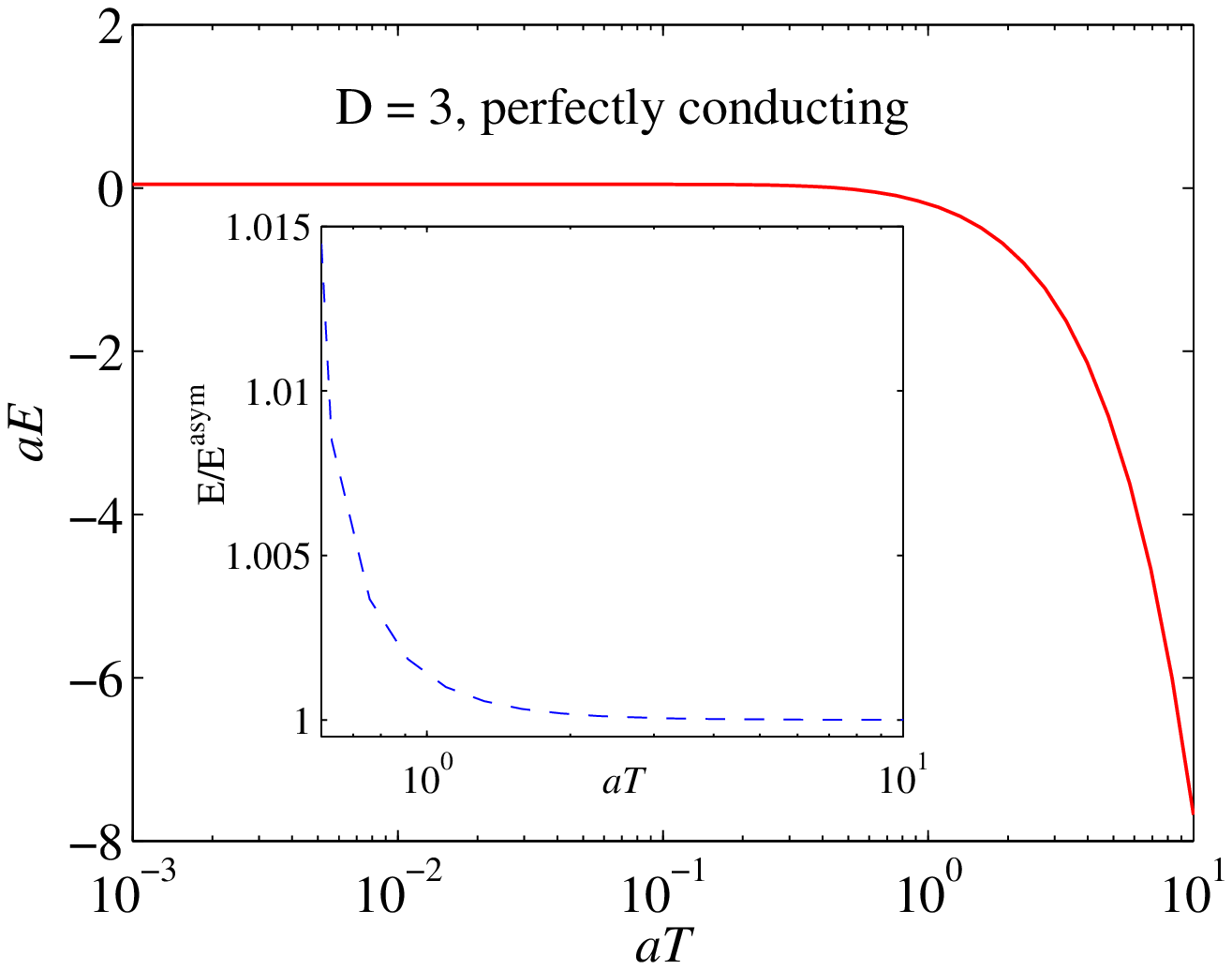}  \epsfxsize=0.43\linewidth \epsffile{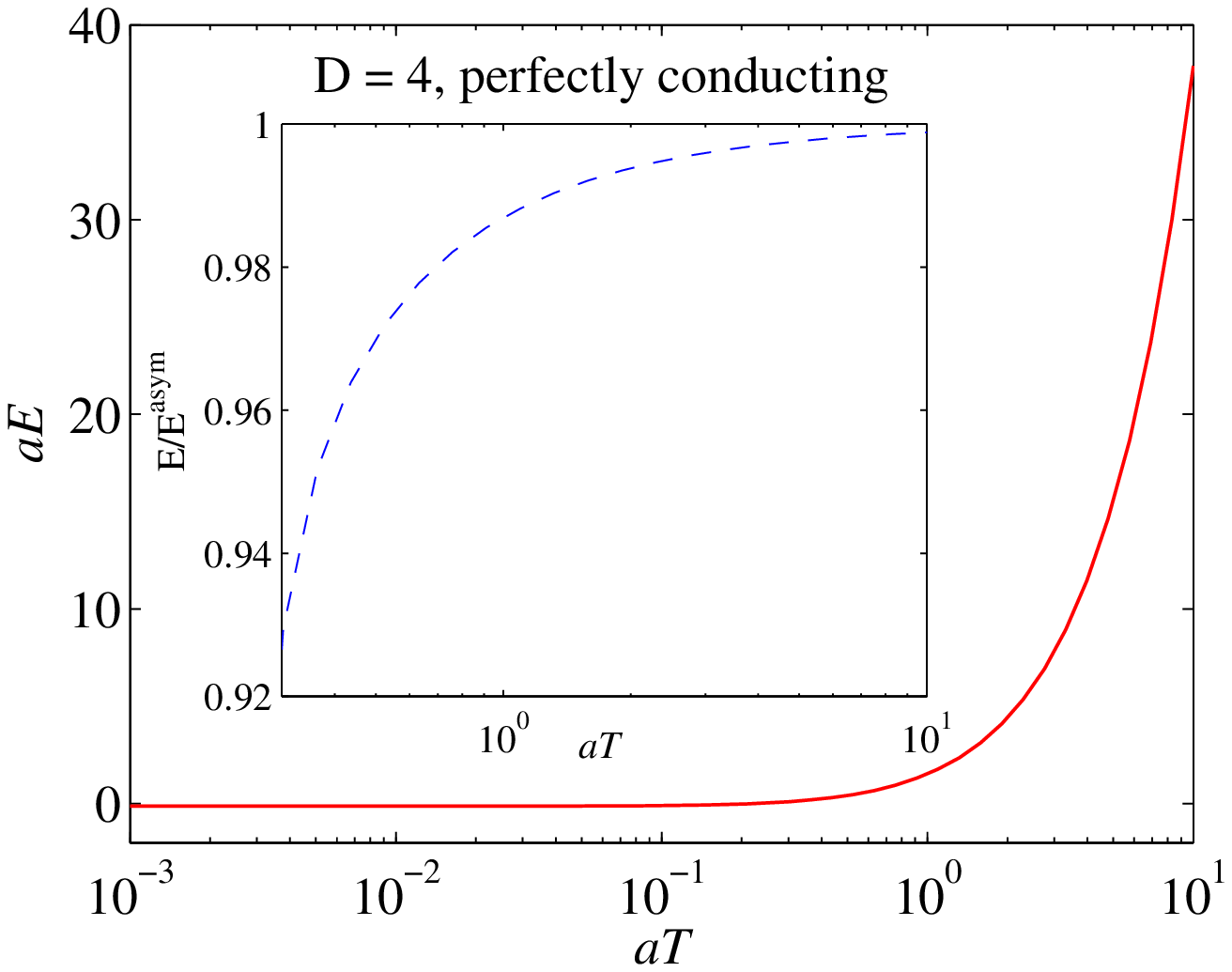}\caption{\label{f5} $aE$  as a function of $aT$ for perfectly conducting boundary conditions when $D=3$ and $D=4$. The inset shows the ratio of $E$ to $E^{\text{asym}}$. }\end{figure}

\begin{figure}
\epsfxsize=0.43\linewidth \epsffile{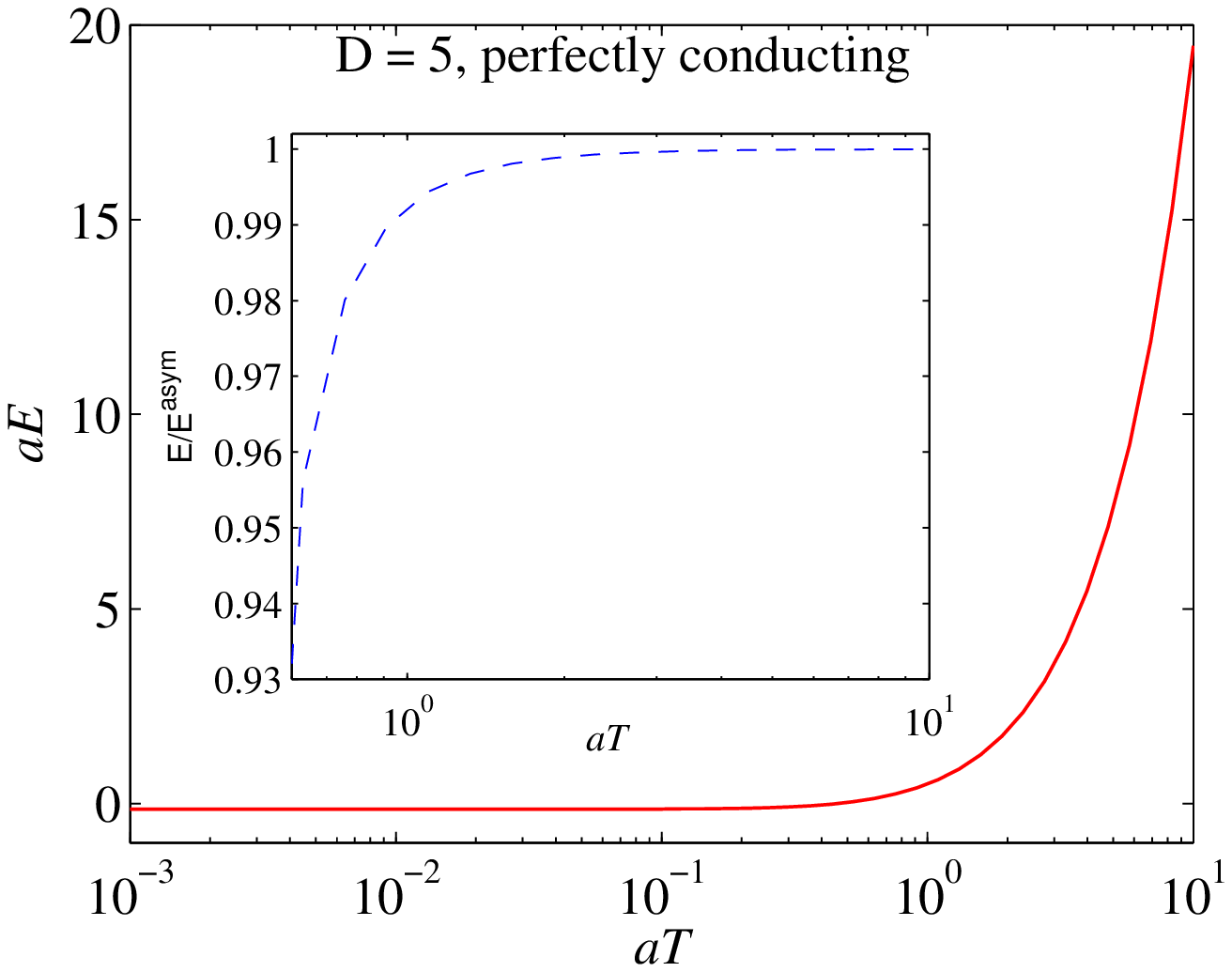}  \epsfxsize=0.43\linewidth \epsffile{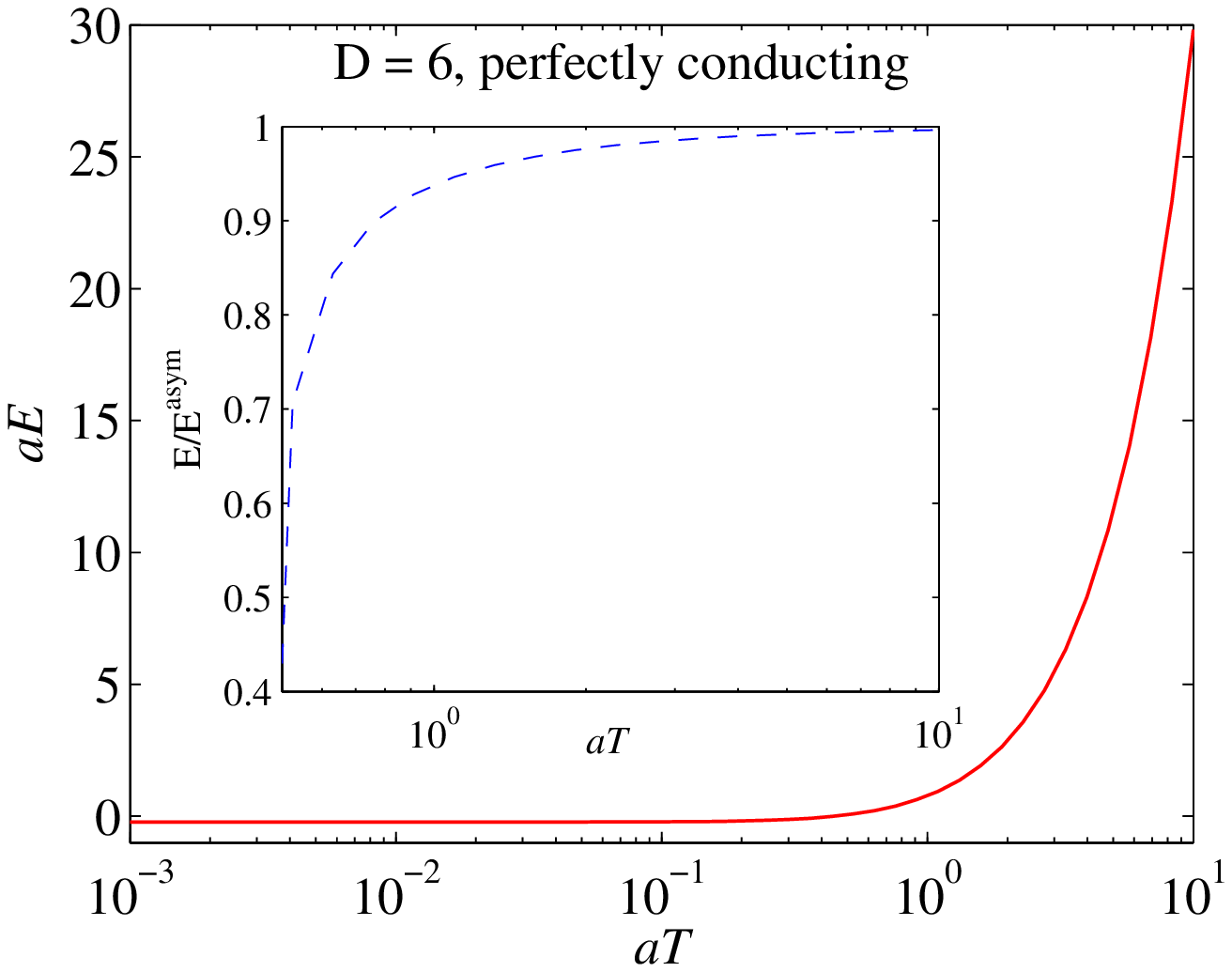}\caption{\label{f6} $aE$  as a function of $aT$ for perfectly conducting boundary conditions when $D=5$ and $D=6$. The inset shows the ratio of $E$ to $E^{\text{asym}}$. }\end{figure}Finally for infinitely permeable boundary condition:
\begin{align*}
E_{\text{Cas}}^{\text{ren}}=&  \frac{1}{4a}\sum_{j=0}^{D-2}\left(x_{D;j}-y_{D;j}\right)\zeta_H\left(-j-1;\tfrac{D}{2}\right) -\frac{1}{2\sqrt{\pi} a}
\sum_{j=0}^{D-2}\left(x_{D;j}-y_{D;j}\right)\sum_{n=0}^{\infty}\sum_{p=1}^{\infty}\left(n+\tfrac{D}{2}\right)^{j+1/2}\sqrt{\frac{2aT}{ p}} K_{ 1/2}\left(\frac{ p}{aT}\left(n+\tfrac{D}{2}\right)\right)\\
&+\frac{1}{2\sqrt{\pi}a}\sum_{i=1}^{\left[\tfrac{D}{2}\right]}\sum_{\substack{j=0\\j\neq 2i-2}}^{D-2}\sum_{k=0}^{2i}\left(x_{D;j}m_{2i,k}\left(\tfrac{4-D}{2}\right)+y_{D;j}d_{2i,k}\right)\frac{\Gamma\left(i+k-\tfrac{1}{2}\right)}
{\Gamma(i+k)}\zeta_H\left(2i-j-1;\tfrac{D}{2}\right)\\&+\frac{1}{4\sqrt{\pi}a}\sum_{i=1}^{\left[\tfrac{D}{2}\right]}
\sum_{k=0}^{2i}\left(x_{D;2i-2}m_{2i,k}\left(\tfrac{4-D}{2}\right)+y_{D;2i-2}d_{2i,k} \right)\frac{\Gamma\left(i+k-\frac{1}{2}\right)}{\Gamma(i+k)}\left(\psi\left(i+k-\tfrac{1}{2}\right)-\psi(1)-2\psi\left(\tfrac{D}{2}\right)\right)\\
&+\frac{2}{\sqrt{\pi}a }\sum_{j=0}^{D-2} \sum_{i=1}^{\left[\tfrac{D}{2}\right]}\sum_{k=0}^{2i}\frac{\left(x_{D;j}m_{2i,k}\left(\tfrac{4-D}{2}\right)+y_{D;j}d_{2i,k} \right)}
{\Gamma(i+k)}\sum_{n=0}^{\infty}\sum_{p=1}^{\infty}
\left(n+\tfrac{D}{2}\right)^{k+j-i+1/2}\left(\frac{ p}{2aT}\right)^{i+k-1/2}K_{i+k-1/2}\left(\frac{  p}{aT}\left(n+\tfrac{D}{2}\right)\right)\\
&-\frac{T}{2}\sum_{l=1}^{\infty}\sum_{p=-\infty}^{\infty}\left(
h_D(l)B_{ \left(\text{R},\tfrac{4-D}{2}\right),\left[\frac{D}{2}\right]}^{ l+\frac{D-2}{2}\prime}(a;0;2\pi p T)+b_D(l)B_{\text{D},\left[\frac{D}{2}\right]}^{ l+\frac{D-2}{2}\prime}(a;0;2\pi p T)\right)\\&-\frac{\hat{c}_{D+1}}{2\sqrt{\pi}}\ln(a\mu)+\frac{1}{\sqrt{\pi}}\sum_{n=0}^{D-1}2^{D-n} \Gamma\left(\frac{D-n+1}{2}\right)\zeta_R(D-n+1) \hat{c}_{n} T^{D-n+1}.
\end{align*}\begin{figure}
\epsfxsize=0.43\linewidth \epsffile{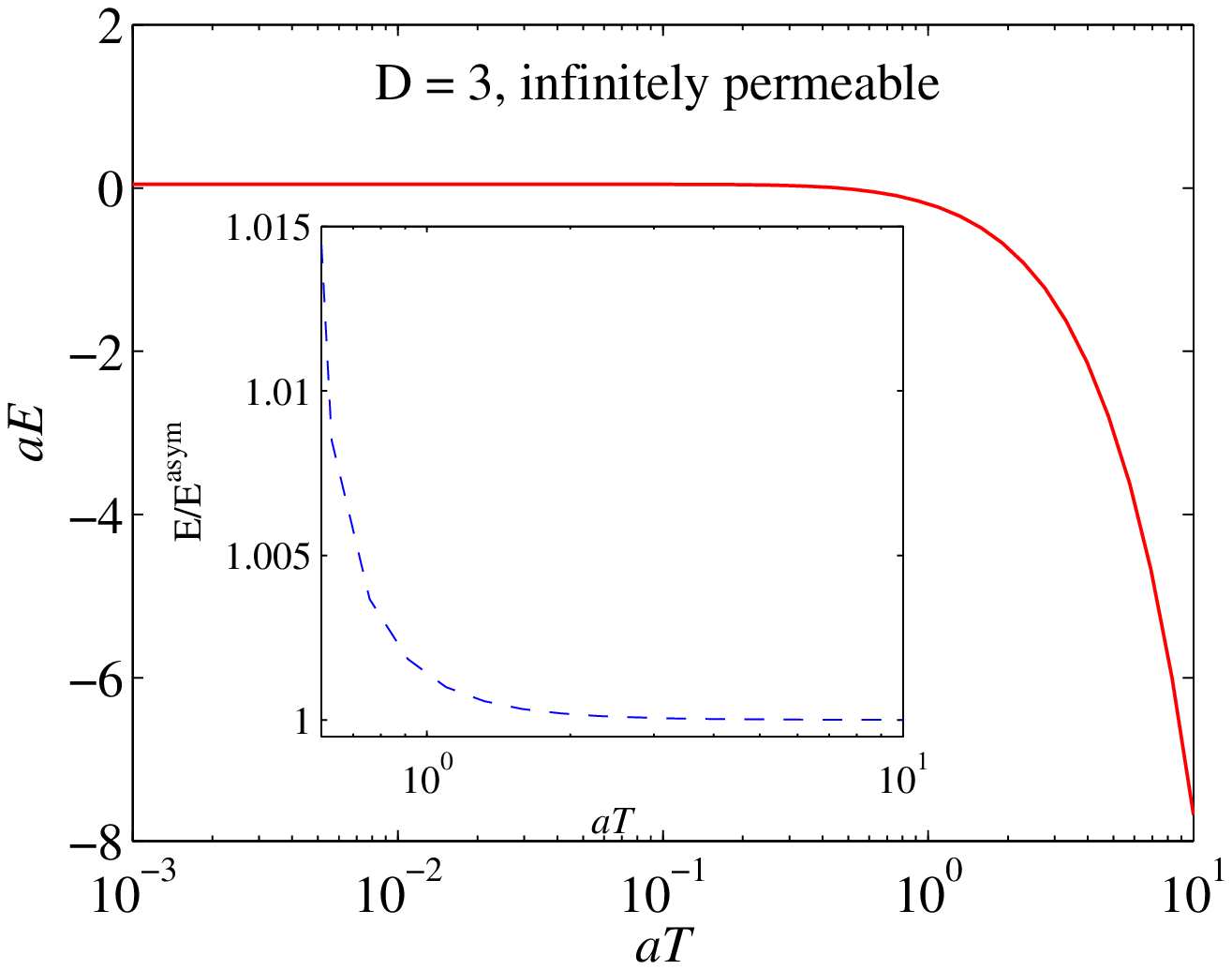}  \epsfxsize=0.43\linewidth \epsffile{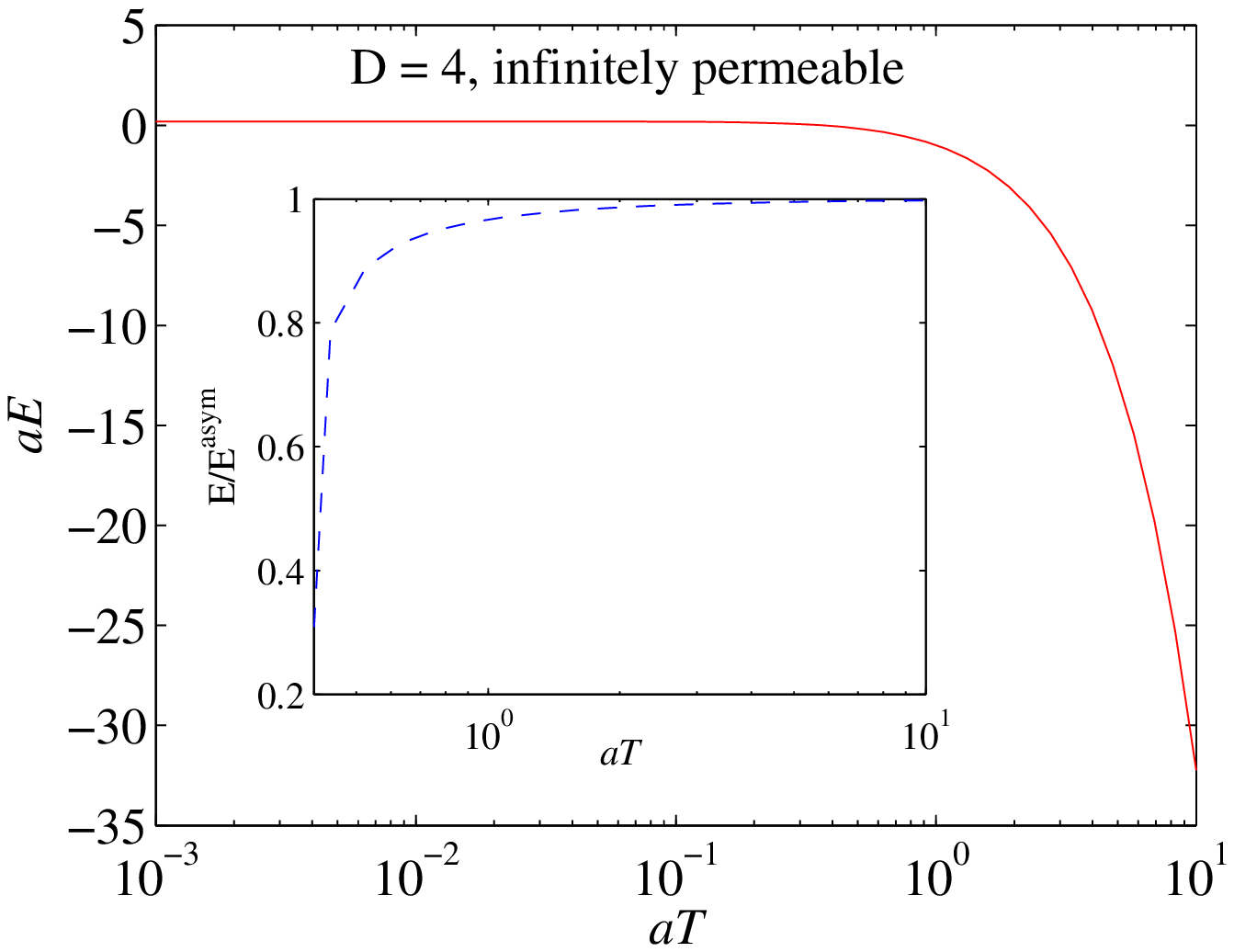}\caption{\label{f7} $aE$  as a function of $aT$ for infinitely permeable boundary conditions when $D=3$ and $D=4$. The inset shows the ratio of $E$ to $E^{\text{asym}}$. }\end{figure}

\begin{figure}
\epsfxsize=0.43\linewidth \epsffile{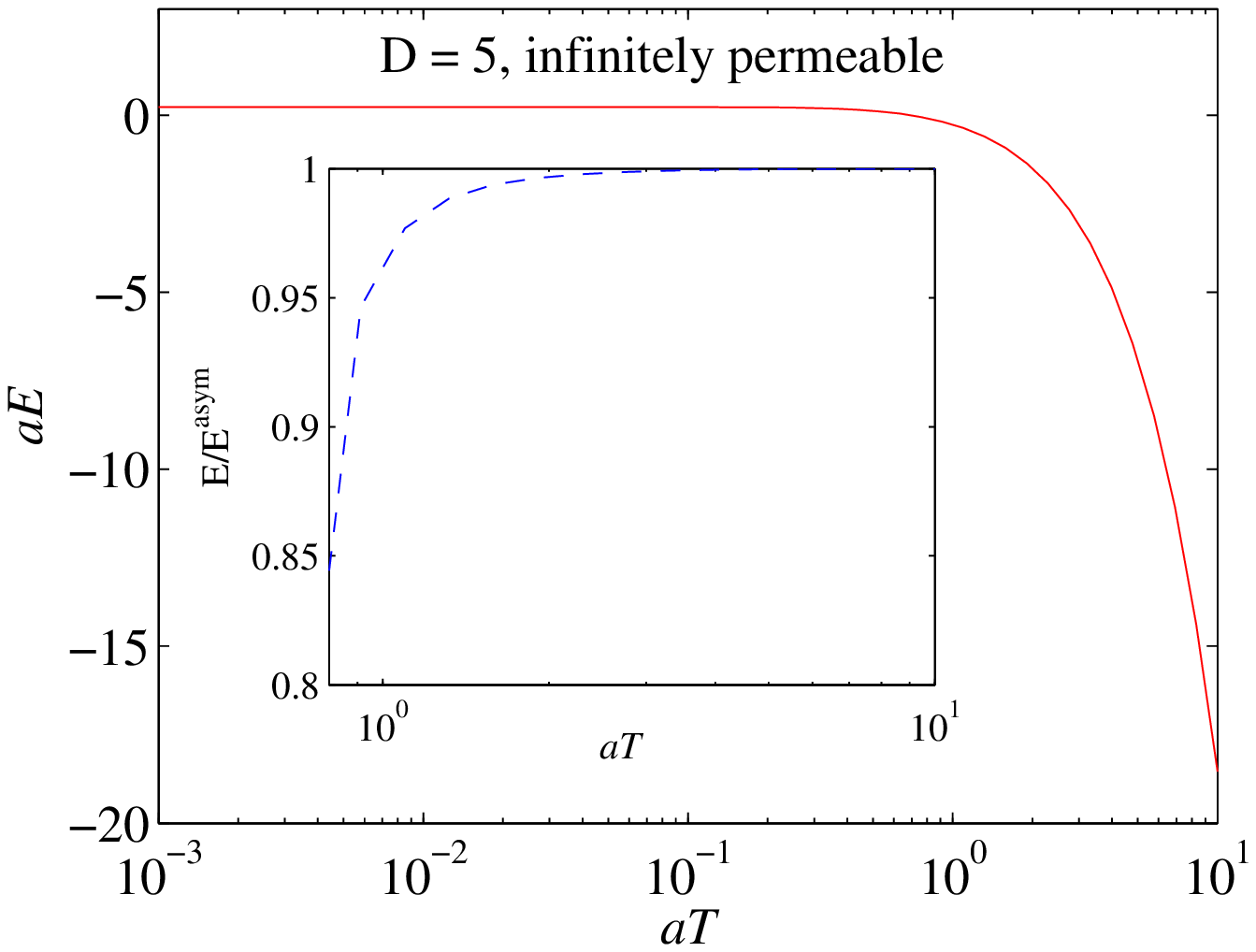}  \epsfxsize=0.43\linewidth \epsffile{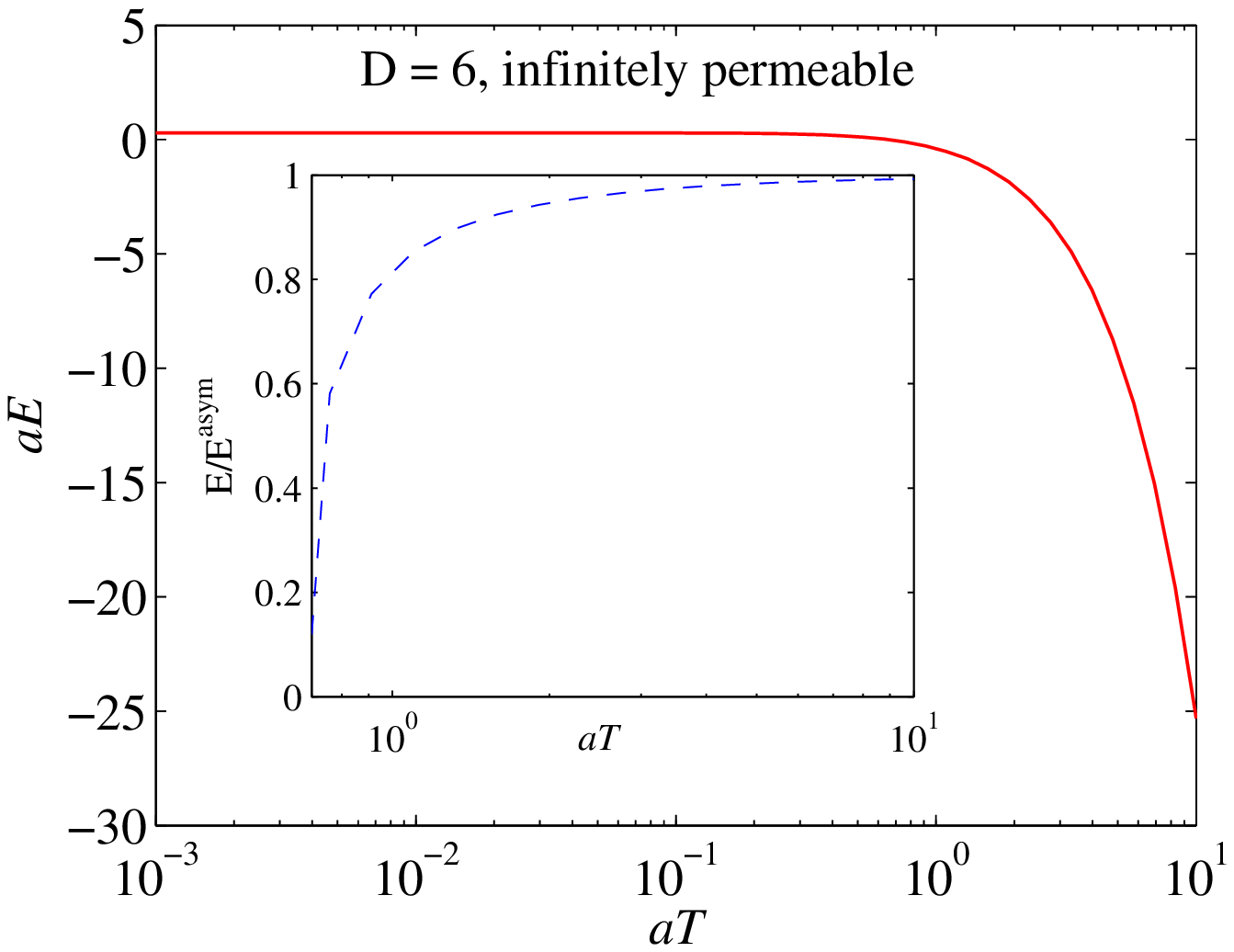}\caption{\label{f8} $aE$  as a function of $aT$ for infinitely permeable boundary conditions when $D=5$ and $D=6$. The inset shows the ratio of $E$ to $E^{\text{asym}}$. }\end{figure}

From these expressions, we see that when $\hat{c}_{D+1}\neq 0$, the renormalized Casimir free energy depend on the normalization constant $\mu$ through the term
$$-\frac{\hat{c}_{D+1}}{2\sqrt{\pi}}\ln(a\mu).$$From \eqref{eq5_7_2}, we find that when $aT\gg 1$,
\begin{equation}
E:=E_{\text{Cas}}^{\text{ren}}+\frac{\hat{c}_{D+1}}{2\sqrt{\pi}}\ln(a\mu)\sim E^{\text{asym}}\end{equation}where
 \begin{equation}E^{\text{asym}}=-T\left(\zeta(a; 0)\ln (aT)+\frac{1}{2}\left[\zeta'(a; 0)-2\zeta(a;0)\ln a\right]\right)
 +\frac{ \psi(1)+\ln (4\pi )+\ln( aT)}{\sqrt{2\pi}}\hat{c}_{D+1}.
\end{equation}
Using the results of Sections \ref{hkc} and \ref{zetaprime}, we listed  $E^{\text{asym}}$ in Tables \ref{t9} and \ref{t10}. For perfectly conducting  spherical shell in $D=3$ dimensions, our result agree with that obtained in \cite{16}.

In Figures \ref{f1}, \ref{f2}, \ref{f3}, \ref{f4}, \ref{f5}, \ref{f6}, \ref{f7} and \ref{f8}, we plot
$$aE=a\left(E_{\text{Cas}}^{\text{ren}}+\frac{\hat{c}_{D+1}}{2\sqrt{\pi}}\ln(a\mu)\right)$$ when $3\leq D\leq 6$, for Dirichlet, Neumann, perfectly conducting and infinitely permeable boundary conditions, as a function of $aT$ for $aT$ between 0.001 and 10,  and compare each the these to  the corresponding $aE_{\text{Cas}}^{\text{asym}}$. From the graphs, it is observed that when $aT> 1$, $E^{\text{asym}}$ gives a good approximation to $E$.

\section{Conclusion}

We have discussed the regularization and renormalization of the Casimir free energy of a spherical shell in $(D+1)$-dimensional Minkowski spacetime. We consider scalar field with Dirichlet and Neumann boundary conditions, and electromagnetic field with perfectly conducting and infinitely permeable boundary conditions. The Casimir free energy  is regularized using  zeta functional method. The  high temperature asymptotic expansion of the regularized Casimir free energy contains terms that are proportional to $\hat{c}_0T^{D+1}, \hat{c}_1T^D, \hat{c}_2 T^{D-1},\ldots, \hat{c}_{D-2}T^3, \hat{c}_{D-1}T^2$, which have to be subtracted away in the renormalization process. Here $\hat{c}_n$ are the heat kernel coefficients. We have derived formulas for these heat kernel coefficients.  When $3\leq D \leq 8$, they are computed explicitly.  For $0\leq n\leq D-1$, it is found that $\hat{c}_n$ is zero if $n$ is even. As is well known, the regularized and renormalized  Casimir energy is unambiguously defined (not depending on a normalization constant) if and only if $\hat{c}_{D+1}=0$, which is the case when $D$ is odd. In the high temperature limit, the leading term of the renormalized Casimir free energy behaves as
$$-\zeta(0)T\ln T-\frac{T}{2}\zeta'(0),$$where $\zeta(s)$ is the corresponding zeta function, and $\zeta(0)=\hat{c}_D$. We have  derived the general expression for $\zeta'(0)$, and computed its explicit value when $3\leq D\leq 8$. This might be of independent interest since it is closely related to a functional determinant. Finally, we derive explicit expressions for the renormalized Casimir free energy. When $3\leq D\leq 6$,     it is computed numerically and the result is shown graphically.

\begin{acknowledgments}\noindent
  This work is supported by the Ministry of Higher Education of Malaysia  under   FRGS grant FRGS/1/2013/ST02/UNIM/02/2. I have benefited from discussions with A. Flachi and K. Kirsten when this work is done.

  \end{acknowledgments}
\appendix
\section{Some computations}\label{a1}
Consider
\begin{align*}
H_D(c)=&  \sum_{l=1}^{\infty}\sum_{j=0}^{D-2}z_{D;j}\nu^{j}B_{\left(\text{R},c\right),N}^{ \nu\prime}(a;0;0),\hspace{2cm}\nu=l+\frac{D-2}{2}\\
=&\sum_{j=0}^{D-2}z_{D;j}\sum_{l=1}^{\infty}\nu^{j}\left(\ln\frac{\nu^2}{\nu^2-c^2}-\sum_{i=1}^N\frac{c^{2i}}{i}\frac{1}{\nu^{2i}}\right)\\
=&-\sum_{j=0}^{D-2}z_{D;j}\sum_{i=\left[\frac{j+3}{2}\right]}^N\frac{c^{2i}}{i}\zeta_H\left(2i-j;\tfrac{D}{2}\right)+Y_D(c),
\end{align*}
where
\begin{align*}
Y_D(c)=\sum_{j=0}^{D-2}z_{D;j}\sum_{l=1}^{\infty}\nu^{j}\left(\ln\frac{\nu^2}{\nu^2-c^2}-\sum_{i=1}^{\left[\frac{j+1}{2}\right]}\frac{c^{2i}}{i}\frac{1}{\nu^{2i}}\right).
\end{align*}
It follows that
$$Y_D(0)=0.$$Taking derivative with of $Y_D(c)$ with respect to $c$,
\begin{align*}
Y_D'(c)=&\sum_{j=0}^{D-2}z_{D;j}\sum_{l=1}^{\infty}\nu^{j}\left(\frac{1}{\nu-c}-\frac{1}{\nu+c}-2\sum_{i=1}^{\left[\frac{j+1}{2}\right]} c^{2i-1} \frac{1}{\nu^{2i}}\right).
\end{align*}
Now,
\begin{align*}
U_{\nu,j}(c)=&\nu^{j}\left(\frac{1}{\nu-c}-\frac{1}{\nu+c}-2\sum_{i=1}^{\left[\frac{j+1}{2}\right]} c^{2i-1} \frac{1}{\nu^{2i}}\right)\\
=&\nu^{j-1}\left(\frac{c}{\nu-c}+\frac{c}{\nu+c}-2\sum_{i=1}^{\left[\frac{j+1}{2}\right]} c^{2i-1} \frac{1}{\nu^{2i-1}}\right)\\
=&\nu^{j-1}c\left(\frac{1}{\nu-c}+\frac{1}{\nu+c}-2\sum_{i=1}^{\left[\frac{j+1}{2}\right]} c^{2i-2} \frac{1}{\nu^{2i-1}}\right)\\
=&\nu^{j-2}c\left(\frac{c}{\nu-c}-\frac{c}{\nu+c}-2\sum_{i=2}^{\left[\frac{j+1}{2}\right]} c^{2i-2} \frac{1}{\nu^{2i-2}}\right)\\
=&\nu^{j-2}c^2\left(\frac{1}{\nu-c}-\frac{1}{\nu+c}-2\sum_{i=1}^{\left[\frac{j-1}{2}\right]} c^{2i-1} \frac{1}{\nu^{2i}}\right)\\
=&\hspace{2cm}\vdots\\
\end{align*}
If $j$ is even,  $U_{\nu,j}(c)$ can be reduced to
\begin{align*}
U_{\nu,j}(c)=&  c^{j}\left(\frac{1}{\nu-c}-\frac{1}{\nu+c}\right).
\end{align*}If $j$ is odd,   $U_{\nu,j}(c)$ can be reduced to
\begin{align*}
U_{\nu,j}(c)=&\nu c^{j-1}\left(\frac{1}{\nu-c}-\frac{1}{\nu+c}-\frac{2c}{\nu^2} \right)\\
=&c^j\left(\frac{1}{\nu-c}-\frac{2}{\nu} +\frac{1}{\nu+c}\right).
\end{align*}
Hence, if $D$ is even,
\begin{align*}
Y_D'(c)=&\sum_{j=0}^{D-2}z_{D;j}c^j \left(-\psi\left(\tfrac{D}{2}-c\right)+ \psi\left(\tfrac{D}{2}+c\right)\right);
\end{align*}whereas if $D$ is odd,
\begin{align*}
Y_D'(c)=&\sum_{j=0}^{D-2}z_{D;j}c^j \left(-\psi\left(\tfrac{D}{2}-c\right)+2\psi\left(\tfrac{D}{2}\right)-\psi\left(\tfrac{D}{2}+c\right)\right).
\end{align*}Here $\psi(z)=d/dz \ln\Gamma(z)$.
Hence, if $D$ is even,
\begin{align*}
Y_D(c)=&\sum_{j=0}^{D-2}z_{D;j}\int_0^cdu\,u^j \left(-\psi\left(\tfrac{D}{2}-u\right)+ \psi\left(\tfrac{D}{2}+u\right)\right)\\
=&-2z_{D;0}\ln\Gamma\left(\tfrac{D}{2}\right)+\sum_{j=0}^{D-2}z_{D;j}\left\{ c^j\ln \left[\Gamma\left(\tfrac{D}{2}+c\right) \Gamma\left(\tfrac{D}{2}-c\right)\right]-
j\int_0^c du\,u^{j-1} \left(\ln\Gamma\left(\tfrac{D}{2}+u\right)+\ln\Gamma\left(\tfrac{D}{2}-u\right) \right)\right\};
\end{align*}
whereas if $D$ is odd,
\begin{align*}
Y_D(c)=&\sum_{j=0}^{D-2}z_{D;j}\left\{2\psi\left(\tfrac{D}{2}\right)\frac{c^{j+1}}{j+1}-\int_0^cdu\,u^j \left(\psi\left(\tfrac{D}{2}-u\right) +\psi\left(\tfrac{D}{2}+u\right)\right)\right\}\\
=&\sum_{j=0}^{D-2}z_{D;j}\left\{2\psi\left(\tfrac{D}{2}\right)\frac{c^{j+1}}{j+1}-c^j\ln\frac{\Gamma\left(\tfrac{D}{2}+c\right)}{\Gamma\left(\tfrac{D}{2}-c\right)}+
j\int_0^c du\,u^{j-1} \left(\ln\Gamma\left(\tfrac{D}{2}+u\right)-\ln\Gamma\left(\tfrac{D}{2}-u\right) \right)\right\}.
\end{align*}
Therefore,
\begin{align*}
Y_D(c)=&-2z_{D;0}\ln\Gamma\left(\tfrac{D}{2}\right)+\sum_{j=0}^{D-2}z_{D;j}\left\{ (1-(-1)^j)\psi\left(\tfrac{D}{2}\right)\frac{c^{j+1}}{j+1} +c^j\left[ \ln\Gamma\left(\tfrac{D}{2}-c\right)+(-1)^j\ln  \Gamma\left(\tfrac{D}{2}+c\right) \right]\right.\\&-\left.
j\int_0^cdu\,u^{j-1} \left((-1)^j\ln\Gamma\left(\tfrac{D}{2}+u\right)+\ln\Gamma\left(\tfrac{D}{2}-u\right) \right)\right\}.
\end{align*}

\end{document}